\documentclass[a4paper]{article}

\usepackage{amsmath, amssymb}
\usepackage{amsfonts}
\usepackage{multicol}
\usepackage{braket}

\usepackage[dvipdfmx]{graphicx}

\usepackage{ascmac}

\usepackage{fancybox}

\usepackage{cite}

\usepackage{bm}

\usepackage{amsthm}

\usepackage{color}

\usepackage{setspace}
\newcommand{\eq}[1]{\begin{equation} #1 \end{equation}}
\newcommand{\eqa}[2]{\begin{equation} #1 \label{#2} \end{equation}}
\newcommand{\balign}[1]{\begin{align} #1 \end{align}}
\newcommand{\bcases}[1]{\begin{cases} #1 \end{cases}}

\newcommand{\ul}{\underline}
\newcommand{\fn}{\footnote}

\newcommand{\figin}[4]
{\begin{figure}[!tb]
\centering
\includegraphics[width= #1]{#2.pdf}
\caption{#3}
\label{f:#4}
\end{figure}}

\newcommand{\todayd}{\the\year/\the\month/\the\day}

\newcommand{\bib}{\bibitem}

\newcommand{\lb}{\label}
\newcommand{\nt}{\notag}
\newcommand{\Tr}{\mathrm{Tr}}

\newcommand{\eref}[1]{Eq.~\eqref{#1}}

\newcommand{\fref}[1]{Supplementary Figure~\ref{f:#1}}

\newcommand{\bel}{\begin{easylist}}
\newcommand{\eel}{\end{easylist}}
\newcommand{\bi}[1]{\begin{itemize} #1 \end{itemize}}
\newcommand{\be}[1]{\begin{enumerate} #1 \end{enumerate}}

\newcommand{\ib}[2]
{\ \begin{itembox}[l]{#1}
#2
\end{itembox} \ }

\def \({\left(}
\def \){\right)}
\def \[{\left[}
\def \]{\right]}

\newcommand{\abs}[1]{\left|#1\right|}

\newcommand{\sumtwo}[2]%
{\mathop{\sum_{#1}}_{#2}}
\newcommand{\sumthree}[3]%
{\mathop{\mathop{\sum_{#1}}_{#2}}_{#3}}
\newcommand{\sumfour}[4]%
{\mathop{\mathop{\mathop{\sum_{#1}}_{#2}}_{#3}}_{#4}} 
\newcommand{\prodtwo}[2]%
{\mathop{\prod_{#1}}_{#2}}
\newcommand{\mintwo}[2]%
{\mathop{\min_{#1}}_{#2}}
\newcommand{\maxtwo}[2]%
{\mathop{\max_{#1}}_{#2}}
\newcommand{\maxthree}[3]%
{\mathop{\mathop{\max_{#1}}_{#2}}_{#3}}
\newcommand{\limtwo}[2]%
{\mathop{\lim_{#1}}_{#2}}
\newcommand{\suptwo}[2]%
{\mathop{\sup_{#1}}_{#2}}
\newcommand{\supthree}[3]%
{\mathop{\mathop{\sup_{#1}}_{#2}}_{#3}}
\newcommand{\supfour}[4]%
{\mathop{\mathop{\mathop{\sup_{#1}}_{#2}}_{#3}}_{#4}} 
\newcommand{\inftwo}[2]%
{\mathop{\inf_{#1}}_{#2}}
\newcommand{\infthree}[3]%
{\mathop{\mathop{\inf_{#1}}_{#2}}_{#3}}
\newcommand{\inffour}[4]%
{\mathop{\mathop{\mathop{\inf_{#1}}_{#2}}_{#3}}_{#4}} 
\newcommand\calA{{\cal A}}
\newcommand\calB{{\cal B}}

\newcommand\calH{{\cal H}}

\newcommand\calS{{\cal S}}
\newcommand\calT{{\cal T}}

\newcommand\calV{{\cal V}}

\newcommand{\bsu}{\boldsymbol{u}}

\newcommand{\bsw}{\boldsymbol{w}}
\newcommand{\bsx}{\boldsymbol{x}}
\newcommand{\bsy}{\boldsymbol{y}}

\newcommand{\bbC}{\mathbb{C}}

\newcommand{\ep}{\varepsilon}

\newcommand{\bA}{\overline{\calA}}
\newcommand{\MG}{M_{\rm G}}
\newcommand{\tMG}{{M}_{\rm G}}
\newcommand{\Heff}{H_{\rm eff}}
\newcommand{\psiin}{\psi_{\rm in}}
\newcommand{\Hin}{\calH^{\rm in}}

\newcommand{\GA}{\Gamma_A}
\newcommand{\tH}{{H}}
\newcommand{\Jy}{J_{\bsy}}
\newcommand{\Jyp}{J_{\bsy'}}
\newcommand{\orho}{\overline{\rho}}

\newcommand{\bl}{b_{\rm l}}
\newcommand{\br}{b_{\rm r}}
\newcommand{\rl}{r_{\rm l}}
\newcommand{\rr}{r_{\rm r}}
\newcommand{\rc}{r_{\rm c}}

\newcommand{\sref}[1]{Sec.~\ref{s:#1}}

\newcommand{\FS}[2]{#1}

\def\rnum#1{\resizebox{0.5em}{\height}{\expandafter{\romannumeral #1}}}
\def\Rnum#1{\resizebox{0.5em}{\height}{\uppercase\expandafter{\romannumeral #1}}}

\makeatletter
  \newcommand{\subsubsubsection}{\@startsection{paragraph}{4}{\z@}%
    {1.0\Cvs \@plus.5\Cdp \@minus.2\Cdp}%
    {.1\Cvs \@plus.3\Cdp}%
    {\reset@font\sffamily\normalsize}
  }
\makeatother
\makeatletter
\def\verbatim@font{\normalfont\fontfamily{txr}\selectfont
\let\do\do@noligs
\verbatim@nolig@list}
\makeatother

\title{\FS{Undecidability of the fate of relaxation in one-dimensional quantum systems}
{Supplementary Note for ``Undecidability in quantum thermalization"}}
\author{Naoto Shiraishi\thanks{Department of physics, Gakushuin university, 1-5-1 Mejiro, Toshima-ku, Tokyo, 171-8588, Japan}
 and Keiji Matsumoto\thanks{Quantum Computation Group, National Institute of Informatics, 2-1-2 Hitotsubashi, Chiyoda-ku, Tokyo 101-8430, Japan}}
\date{}

\FS{}{

\setcounter{equation}{0}

}

\begin{document}

\maketitle

\FS{
\begin{abstract}
We investigate the relaxation dynamics in an isolated quantum many-body system.
The stationary value of an observable after relaxation is a topic of research in quantum thermalization since thermalization is a relaxation phenomenon where this stationary value coincides with the equilibrium value.
Therefore, computing the stationary value in quantum many-body systems is regarded as an important problem.
We, however, prove that the stationary value in quantum many-body systems is incomputable.
More precisely, we show that whether the stationary value is in the vicinity of a given value or not is an undecidable problem.
Our undecidable result is still satisfied when we restrict our system to a one-dimensional shift-invariant system with nearest-neighbor interaction, our initial state to a product state of a state on a single site, and our observable to a shift-sum of a one-body observable.
This result clearly shows that there is no general theorem or procedure to decide the presence or absence of thermalization in a given quantum many-body system.
\end{abstract}
}{}

\tableofcontents

\section{Introduction}

\FS{
Thermalization in quantum many-body systems has attracted the interest of researchers in various research fields.
In these fields, we investigate what determines the presence or absence of thermalization and how to understand thermalizing and non-thermalizing phenomena.
This problem is an old longstanding problem, which has already been discussed by Boltzmann~\cite{Bol} and von Neumann~\cite{Neu}.
The recent development of experimental techniques on cold atoms pushes this old problem toward a new step, and thermalization of isolated quantum many-body systems has been minutely examined in laboratory~\cite{KWW, Tro, Gri, Lan, Kau}.
Elaborated experiments have revealed some unexpected behaviors, including quantum many-body scars~\cite{Ber}, which has recently been studied intensively~\cite{Tur18, LM18, Ho, Shi19, SAP}.
In the theoretical approach, various suggestive notions, including typicality of states~\cite{Llo, PSW, Gold, Sug, Rei07, Tas16} and dynamics~\cite{BG09, GHT, Rei16, DR20}, eigenstate thermalization hypothesis~\cite{Deu, Sre, HZB, Tas98, Rig08, Bir} and its violation~\cite{SM, MS, Ber17, Mou, Shi17}, weak-version of eigenstate thermalization hypothesis~\cite{Bir, Mor16}, the effective dimension~\cite{Rei08, LPSW, SF12, FBC16, Tas16}, the quantum many-body localization~\cite{BAA, PH, HSS, SPA, Fri15, NH15, Imb}, generalized Gibbs ensemble~\cite{Caz, RDYO, Ill, EF} and many other important findings~\cite{GME, Kim, Pin, S18} have been raised, and some aspects of thermalization have been well clarified (see a review paper~\cite{GE16}).
However, the full understanding of thermalization is still far from our present position.
In particular, one of the central problems, determining whether a given system thermalizes or not, has still been left unsolved.

In this paper, we consider this problem by a completely different approach from conventional ones.
We employ the ideas from theoretical computer science and quantum information and prove that the above problem of thermalization is undecidable.
More precisely, we prove that the expectation value of an observable $A$ after relaxation with a given Hamiltonian $H$ of a one-dimensional system is undecidable.
Our result of undecidability is still valid if we restrict the class of the observable $A$ as a shift sum of a given one-body observable, the Hamiltonian $H$ as a shift-invariant nearest-neighbor interaction, and the initial state as a product state of identical states on a single site (except the first site).
This result directly implies that there is no general procedure to decide the presence or absence of thermalization, obviously no general theorem.

Undecidability sometimes appears in physics.
Examples are dynamical systems~\cite{Moo}, repeated quantum measurement~\cite{EMG}, and the spectral gap in quantum many-body systems~\cite{Cub}, and our finding tells that quantum thermalization stands in this line.
The following two tools inspire our proof: the reduction to the halting problem of Turing machine~\cite{Tur, MMbook} and Feynman-Kitaev type quantum emulation of classical machines~\cite{Fey, Kit}.
By adopting several technical ideas, we successfully construct a quantum many-body system where the halting problem of Turing machine determines the destination after relaxation.
}{
This Supplementary Note aims to provide the full proof of the theorem and lemma in the main text (which is called Theorem 1b, Theorem 2, and Lemma 1 in this Supplementary Note).
The following two tools inspire our proof: the reduction to the halting problem of Turing machines~\cite{Tur, MMbook} and Feynman-Kitaev type quantum emulation of classical machines~\cite{Fey, Kit}.
Combining these ideas properly, we successfully construct a quantum many-body system where the halting problem of Turing machines determines the destination after relaxation.
}

This \FS{paper}{Supplementary Note} is organized as follows:
In \sref{back} and \sref{back-computer}, we provide a pedagogical review of quantum thermalization and theoretical computer science.
In \sref{result}, we state three main theorems and a technical lemma.
The latter lemma is the most important result from a theoretical aspect.
In \sref{pf-thm-lemma}, we derive two theorems, undecidability of relaxation, from the technical lemma.
Most of the remainder of this \FS{paper}{Supplementary Note} is devoted to proving this lemma.
In \sref{strategy}, we briefly sketch the proof strategy.
In \sref{pf1}, we introduce a classical universal reversible Turing machine, which is responsible for the halting problem.
In \sref{pf2}, we construct the Hamiltonian of the quantum system emulating the classical dynamics of the Turing machine.
Since the details of dynamics of our quantum system are a little complicated, in \sref{pf3} we introduce an analogous setting, which is easier to treat, and solve the dynamics.
In \sref{pf4}, we go back to the original setting and construct the initial state with which the expectation value after relaxation is indeed undecidable.
In \sref{proof-thermal}, we prove the remaining theorem, undecidability of thermalization, by extending the proof techniques for the previous lemma.
In \sref{remark}, we briefly comment on what happens if we numerically simulate the constructed system.

\FS{
This paper also serves as the supplementary note for the short letter~\cite{letter}.
Two main theorems and a key lemma in Ref.~\cite{letter} correspond to Theorem 1b, Theorem 2, and Lemma 1 in this paper, respectively.
}{}


\section{Background of quantum thermalization}\lb{s:back}

Before going to our main result, we first summarize the problem of thermalization and theoretical computer science in this and next sections for readers who are not familiar with these topics.
If a reader is familiar with them, one can skip this and the next sections.

In the research field of quantum thermalization, we mainly consider whether a quantum many-body system with a Hamiltonian $H$ at the initial state $\ket{\psi}$ thermalizes (with respect to an observable $\calA$) or not.
A state $\ket{\phi}$ is called {\it thermal} with respect to an observable $\calA $ if its expectation value of $\calA $ is close to its equilibrium value:
\eq{
\braket{\phi|\calA |\phi}\simeq \Tr[\calA \rho^{\rm MC}],
}
where $\rho^{\rm MC}$ is a microcanonical state with energy $\braket{\phi|H|\phi}$.
The symbol $\simeq$ means that both-hand sides coincide in the thermodynamic limit (i.e., If $\calA =O(1)$, $\calA \simeq \calB $ means $\lim_{V\to \infty}(\calA -\calB )=0$, and if $\calA =O(V)$, $\calA \simeq \calB $ means $\lim_{V\to \infty}\( \frac \calA V- \frac \calB V\) =0$).
We call that an initial state $\ket{\psi}$ under the Hamiltonian $H$ {\it thermalizes} with respect to $\calA =O(1)$
\fn{
It is easy to extend the definition to the case of $\calA=O(V)$.
}
if $\ket{\psi(t)}:=e^{-iHt}\ket{\psi}$ is thermal with respect to $\calA $ for almost all $t$, that is, 
\eq{
\lim_{V\to \infty} \lim_{T\to \infty}\frac1T \int_0^T dt \chi\{ \abs{\braket{\psi(t)|\calA |\psi(t)}- \Tr[\calA \rho^{\rm MC}]}<\ep \} = 1
}
is satisfied for any $\ep>0$.
Here $\chi\{\cdot \}$ is the indicator function which takes 1 (resp. 0) if the statement inside the bracket is true (resp. false).
In the above case, $\chi\{ \cdot \}$ takes one if $\abs{\braket{\psi(t)|\calA |\psi(t)}- \Tr[\calA \rho^{\rm MC}]}<\ep$ is satisfied at the time $t$, and takes zero otherwise.
Note that due to the quantum recurrence theorem~\cite{BL57}, for any $T'$ there exists $\tau>T'$ such that the state at time $\tau$, $\ket{\psi(\tau)}$, and the initial state $\ket{\psi(0)}$ is arbitrarily close.
Our definition of thermalization allows recurrence, while recurrence time should become extremely long in a large system.

Let $\bar{\calA }:=\lim_{T\to \infty}\frac1T \int_0^T dt \braket{\psi(t)|\calA |\psi(t)}$ be the long-time average of $\calA $.
Then, an initial state thermalizes if the following two conditions are satisfied:
\bi{
\item (Relaxation): The time-series fluctuation around the long-time average $\bar{\calA }$ converges to zero:
\eq{
\lim_{T\to \infty}\frac1T \int_0^T dt (\braket{\psi(t)|\calA |\psi(t)}-\bar{\calA })^2\simeq 0.
}
\item (Convergence to the equilibrium value): The long-time average $\bar{\calA }$ converges to the equilibrium value $\Tr[\calA \rho^{\rm MC}]$:
\eq{
\bA\simeq \Tr[\calA\rho^{\rm MC}].
}
}
The former condition, relaxation, is proven for initial states under some condition (a diverging effective dimension)~\cite{Rei08, LPSW, SF12} and this condition is shown to be fulfilled analytically in physically plausible initial states~\cite{FBC16}.
Thus, the remaining hard task is to handle the latter condition, convergence to the equilibrium value.
This is why various concepts and arguments raised in the field of quantum thermalization concern the long-time average and the equilibrium value.

The main goal of this \FS{paper}{Supplementary Note} is to prove the incomputability of the value of the long-time average $\bA$.
In other words, we have no general procedure to detect the behavior of quantum many-body systems after relaxation.

\section{Background of theoretical computer science}\lb{s:back-computer}

We here briefly review some basic notions of theoretical computer science; the Turing machines, the Church-Turing thesis, decision problems, and the halting problem of Turing machines.
Detailed explanations can be seen in e.g., the textbook of Moore and Mertens~\cite{MMbook}.

\figin{7cm}{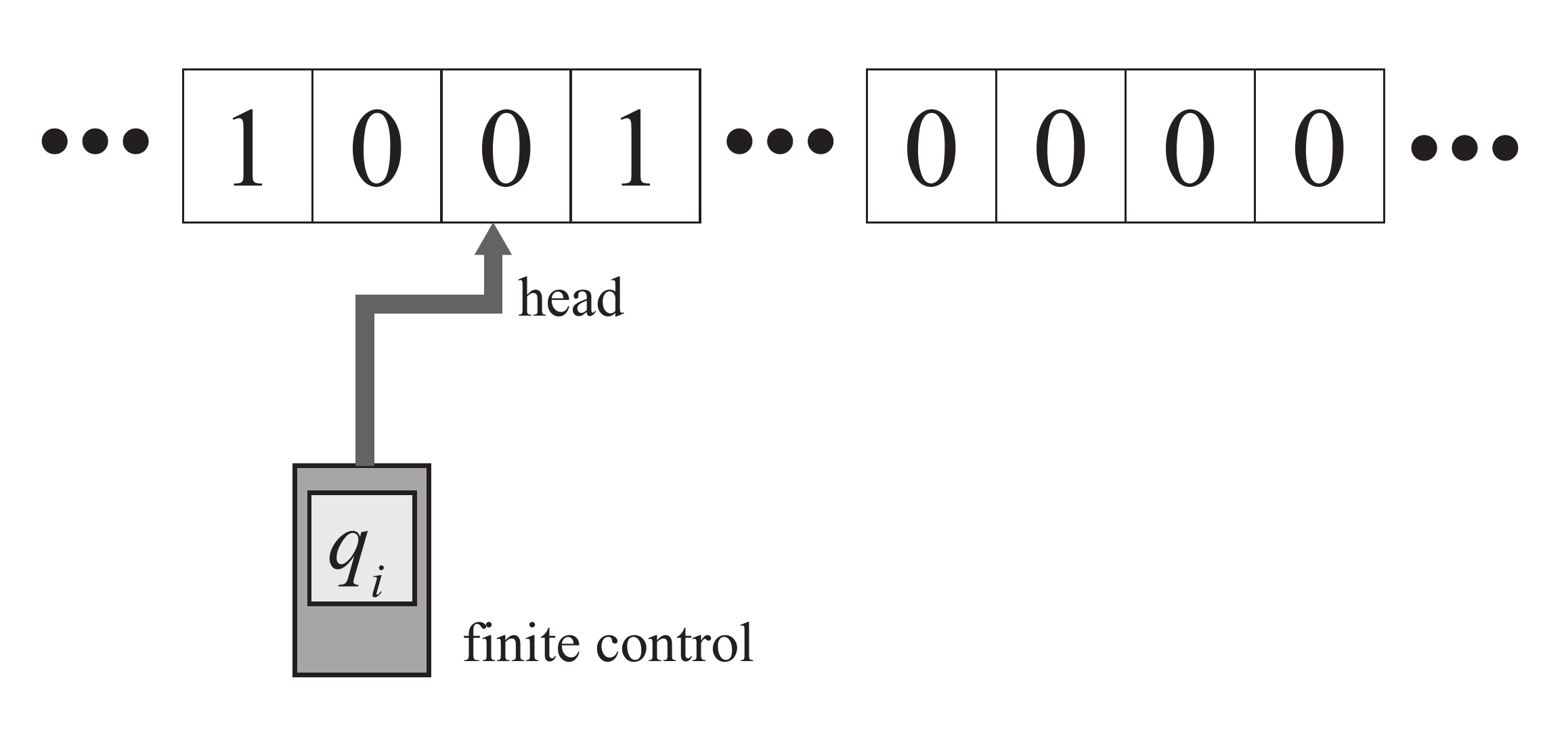}{
Schematic of a Turing machine (TM).
The finite control with the internal state $q_i$ reads the cell with its symbol $0$.
}{TM-explain}

We employ the description with Turing machines (TM) as computation.
The TM is a very simple computation system which consists of a one-dimensional {\it tape} (an infinite line of cells) filled with symbols and a {\it finite control} with a {\it head} and its own {\it internal state} (see \fref{TM-explain}).
The head reads a single cell on the tape, and can rewrite the symbol of the cell, and can move left or right one cell.
At the same time, the internal state of the finite control may change.
The {\it transition function} specify the rule of dynamics in the following manner as an example:

\begin{quote}
If the internal state of the finite control is $q_i$ and the symbol in the cell read by the head is $g_m$, then the machine changes the internal state to $q_j$, rewrites the cell to $g_n$, and moves the head left.
\end{quote}

A TM is defined as a set of tape alphabets, a set of internal states, and a transition function\fn{
Precisely, we also require that the tape alphabet should contain a blank cell, the set of internal state should contain a start state and a halting state
}.
One may feel that TMs are very primitive and too simple.
However, there exist universal Turing machines which can implement all possible TMs, and these universal Turing machines can implement almost all computational tasks in our world (e.g., computation by C++ and Python).
Thus, it is declared that computational functions are those computable by TMs, which is known as the Church-Turing thesis.
By accepting the Church-Turing thesis, our computational ability is equivalent to that of a TM, and therefore we can safely restrict our attention to a universal Turing machine.

\bigskip

We next explain {\it decision problems}.
In decision problems, a set of infinite inputs with a Yes/No assignment is given, and our task is to answer Yes/No correctly for all inputs.
In the case of the primary test, for example, inputs are natural numbers, and we need to answer Yes if and only if the input is a prime number.
If there exists a TM which outputs 1 with any input assigned to Yes and outputs 0 with any input assigned to No, this decision problem is called {\it decidable}.
If a decision problem is not decidable (i.e., no TM answers this problem correctly), this decision problem is called {\it undecidable}.
We note that the speed of computation does not matter to decidability.
In fact, combinatorial optimization problems, which are regarded as hard tasks in practice, are categorized as decidable problems because we can solve them by brute force methods.
We also note that undecidable problems must accompany infinite possible inputs because there always exists a TM whose output coincides with the correct answer for finite inputs accidentally.
This is one of the reasons why standard setups of theoretical computer science usually accompany infinite inputs.

A famous undecidable problem is {\it the halting problem of TMs}.
In this decision problem, inputs are input codes for a fixed universal Turing machine\fn{
Recall that a single universal Turing machine emulates any TM with any input.
}, 
and we need to answer whether this TM halts at some time or moves forever.
It is proved that this decision problem is undecidable.
In our proof of the undecidability of thermalization, we reduce our problem to the halting problem of TMs.

We sometimes treat decision problems {\it with promise}.
In the case with promise, the inputs are restricted to a subclass of possible inputs where this promise is satisfied.
In other words, we allow not to answer or to answer incorrectly if the input does not satisfy this promise.
We again take the primary test as an example.
Suppose a promise that the input is a prime or a product of two primes.
Then, if the input is a product of three primes, we may incorrectly answer Yes (a TM may output 1) for this input.

\section{Setup and main results}\lb{s:result}

We first state the undecidability of {\it relaxation} and then state the undecidability of {\it thermalization} in this section.
All the essential ideas for the proof of the undecidability of thermalization have already appeared in that of relaxation.
Therefore, in this \FS{paper}{Supplementary Note} we first treat that of relaxation, which spends most of this \FS{paper}{Supplementary Note}, and then briefly discuss some additional ideas to extend the result of relaxation to thermalization.

Consider a one-dimensional chain of $d$-level quantum systems with the periodic boundary condition, whose underlying local Hilbert space is denoted by $\calH$.
Suppose that the Hamiltonian of this chain $H$ is shift-invariant and contains only 1-body terms and nearest-neighbor 2-body interactions.
We set the length of the system as $L$.
Let $A$ be a non-negative observable of $\calH$, and $\calA_L$ be the normalized spatial average of $A$'s:
\eq{
\calA_L:=\frac{1}{L} \sum_{i=1}^{L}A_i,
}
where $A_i$ is the operator $A$ acting on the site $i$.
Our interest is a long-time average of $\calA_L$:
\eq{
\bA(H,\rho):=\lim_{L\to \infty}\lim_{T\to \infty}\frac{1}{LT}\int_0^T \Tr[e^{-iHt}\rho_L e^{iHt}\calA_L] dt,
}
where $\rho_L$ is the initial state of the system\fn{
More precisely, we first introduce a state $\rho$ with infinite length, and define $\rho_L$ as the restriction of the state $\rho$ to the first $L$ consecutive sites.
}
with length\fn{\lb{footnote-limit}
We note that how the system size increases (e.g., $1\leq i\leq L$ or $-L/2+1\leq i\leq L/2$) does not matter to our result of undecidability because this system has the periodic boundary condition and almost uniform initial states in the form of $\ket{\phi_0}\ket{\phi_1}^{\otimes L-1}$ suffices to show the undecidability.
}
$L$.
Whenever without confusion, the dependency of $\bA$ on $H$ and $\rho$ is dropped.

We argue, roughly, that $\bA$ is incomputable\fn{
To make such an assertion appropriately, some vocabulary from computer science is necessary: 
We say the operator $A$ is computable if a Turing machine can compute any component (with respect to some standard basis) of $A$ with any specified accuracy (So the inputs of a Turing machine are the indices of the components and the integer specifying the accuracy).
Also, the Hamiltonian and the density operator over the one-dimensional chain are computable if their restriction to the finite system size ($L$) is computable. 
In our case, the Hamiltonian is computable if and only if its 1- and 2- body terms are computable.

Whenever we say an operator $A$ is an input to the problem, we mean that the bit string describing $A$ is given to a Turing machine as the input. 
Here, the map of the bit string to an arbitrary component should be computable up to any given accuracy. 
When the input is an observable or a state over $\calH^{\otimes \infty}$, the bit string description is given again.
The computability of the map of the bit string to the observable or the state is defined considering restriction to the finite-size system.
}.
Our statement still holds even in the case that the Hamiltonian is shift-invariant and nearest-neighbor interaction, and the initial state is an almost uniform product state in the following form:
\eqa{
\rho_L=\ket{\phi_0}\ket{\phi_1}^{\otimes L-1}.
}{ini-form}
To state our claim in a rigorous manner, we define two decision problems with a promise; STA (state time average) and HTA (Hamiltonian time average).

\begin{quote}
{\bf [STA]}: 
The dimension of the local Hilbert space $d=\dim \calH$, a Hamiltonian $H$, and a one-body observable $A$ are fixed.
These two, $H$ and $A$, are parameters of the problem\fn{
We also require that $H$ and $A$ are computable.
}.

\ul{Input}: a density operator $\rho$.

\ul{Promise}: Either $\bA\in [c+\ep_1, c-\ep_1]$ or $\bA\notin [c+\ep_2, c-\ep_2]$ with $0<\ep_1<\ep_2$ holds.

\ul{Decision problem}: Decide which of the above two holds.
\end{quote}

\bigskip

\begin{quote}
{\bf [HTA]}: 
The dimension of the local Hilbert space $d=\dim \calH$, a one-body observable $A$, and an initial state $\rho$ are fixed.
These two, $A$ and $\rho$, are parameters of the problem\fn{
We also require that $A$ and $\rho$ are computable.
}.

\ul{Input}: a Hamiltonian $H$.

\ul{Promise}: Either $\bA\in [c+\ep_1, c-\ep_1]$ or $\bA\notin [c+\ep_2, c-\ep_2]$ with $0<\ep_1<\ep_2$ holds.

\ul{Decision problem}: Decide which of the above two holds.
\end{quote}

We now state our main theorems, which claim that these decision problems are undecidable.
Below, the necessary dimension $d_0$ is a fixed number, whose rough estimation is presented in \sref{dim} as $d_0\simeq 120$.

We first state the undecidability of relaxation in the form of STA:

\ib{Theorem 1a}{
Fix the local dimension $d\geq d_0$.
Then, there exists a family of shift-invariant Hamiltonians $H$ with 1-body terms and 2-body nearest-neighbor interaction terms having the following properties:

Let $A$ be an arbitrary observable on a single site and $\ep>0$ be an arbitrary error margin.
We put no requirement on the observable.

Then, for any $M>1$ there exist an operator $A'$ with $\|A-A'\|\leq \ep$, a value $A^*$, and proper errors $\ep_1$ and $\ep_2$ with $\ep_2\leq \ep$ and $\ep_2/\ep_1\geq M$ such that the STA with the aforementioned $A'$, $H$, $c=A^*$, $\ep_2$ and $\ep_1$ is undecidable.
This remains valid even if the initial state $\rho$ (inputs for STA) is restricted to a product state of a pure state of a single site in the form of \eref{ini-form}, where $\ket{\phi_0}$ and $\ket{\phi_1}$ are orthogonal to each other.
}

We here clarify the dependency of quantities:
The system Hamiltonian $H$ is fixed independent of the observable $A$ and a small parameter $\ep$.
In contrast, the target value $A^*$ and two parameters for errors $\ep_1$ and $\ep_2$ depend on $H$, $A$, and the parameters $\ep$ and $M$.
The task of STA is to determine $\bA\in [c+\ep_1, c-\ep_1]$ or $\bA\notin [c+\ep_2, c-\ep_2]$ with $c=A^*$ for all inputs $\rho$ with the form \eqref{ini-form}, and thus to show undecidability we pick up a family of {\it bad} inputs depending on $A$, $\ep_1$, and $\ep_2$.

We slightly modified the observable from $A$ to $A'$ in order to avoid several unwanted cases that the observable $A$ is an identity operator and that the basis determined by the Hamiltonian $H$ and the eigenbasis of $A$ are accidentally aligned in an unwanted direction.

We next state the undecidability of relaxation in the form of HTA.
The next theorem on HTA, Theorem 1b, is stronger than Theorem 1a in that two of three main quantities, the initial state, the observable, and the Hamiltonian, are arbitrary.

\ib{Theorem 1b}{
Fix the local dimension $d\geq d_0$.
Let $A$ be an arbitrary observable on a single site and $\rho$ be an arbitrary pure state in the form of \eref{ini-form}.
We require that there exists a state $\ket{\phi_2}$ orthogonal to $\ket{\phi_0}$ and $\ket{\phi_1}$ such that $\braket{\phi_2|A|\phi_2}\neq \braket{\phi_1|A|\phi_1}$.

Then, for any large $M>1$, there exist proper $\ep_2$, $\ep_1$ with $\ep_2/\ep_1\geq M$ and $A^*$ such that the HTA for $A$ and $\rho$ with $c=A^*$, $\ep_2$ and $\ep_1$, is undecidable.
This remains valid even if the Hamiltonian $H$ (inputs of HTA) is restricted to a shift-invariant Hamiltonian $H$ with 1-body terms and 2-body nearest-neighbor interaction terms.
}

Here, we put the assumption on $\ket{\phi_2}$ in order to exclude the case that $A$ is close to an identity operator.

We again clarify the dependency of quantities:
Both the observable $A$ and the initial state $\rho$ are arbitrary (as far as the condition for $\ket{\phi_2}$ is satisfied).
Two parameters for the error, $\ep_1$ and $\ep_2$, depend on the choice of $A$ and $\rho$.
In addition, the target value $A^*$ also depends on the choice of $A$ and $\rho$.
The task of HTA is to determine $\bA\in [c+\ep_1, c-\ep_1]$ or $\bA\notin [c+\ep_2, c-\ep_2]$ with $c=A^*$ for all inputs $H$ in the form of shift-invariant and nearest-neighbor interaction, and thus to show undecidability we pick up a family of {\it bad} inputs depending on $A$, $\rho$, $\ep_1$, and $\ep_2$.

\bigskip

These two theorems are, in fact, different rewritings of the following technical lemma.
This lemma claims that there exists a shift-invariant one-dimensional Hamiltonian which decodes the input code from the Hamiltonian itself and emulates the dynamics of a universal reversible Turing machine (URTM) properly.
Note that in this lemma, $\bsu$ takes all possible binary bit strings: $\bsu\in \cup_{n=1}^\infty \{ 0,1\}^n$.

\ib{Lemma 1}{
Fix the dimension of the local Hilbert space at $d\geq d_0$.
A complete orthonormal system (CONS) of the local Hilbert space $\{ \ket{e_i}\}_{i=0}^{d-1}$ and an observable $A$ over $\calH$ with $\braket{e_1|A|e_1}=0$ and $\braket{e_2|A|e_2}>0$ are given arbitrarily.
A universal reversible Turing machine (URTM) on a single tape is given arbitrarily.
Then, for any $\eta>0$ there exist a shift-invariant Hamiltonian $H$ (which depends on the CONS) and a set of computable unitary operators $\{ V_{\bsu}\}$ over $\calH$ (which depends on the CONS, $A$, and $\eta>0$)  with the following properties:
\bi{
\item $H$ consists of 1-body terms and 2-body nearest-neighbor interaction terms.
\item For any $\bsu$, $V_{\bsu}\ket{e_0}=\ket{e_0}$ is satisfied.
\item By setting the initial state as
\eq{
\ket{\psi_{V,L}}:=V_{\bsu}\ket{e_0}\otimes (V_{\bsu}\ket{e_1})^{\otimes L-1},
}
then with defining $\calV:=(V_{\bsu})^{\otimes L}$
\eqa{
\min \{ \bA(H,\psi_V), \overline{\calV\calA \calV^\dagger}(H, \psi_V)\} \geq \( \frac14 -\eta\) \braket{e_2|A|e_2}
}{lemma-halt}
is satisfied if and only if the URTM halts with the input $\bsu$, and 
\eqa{
\max \{ \bA(H,\psi_V), \overline{\calV\calA \calV^\dagger}(H, \psi_V)\} \leq \eta
}{lemma-nonhalt}
is satisfied if and only if the URTM does not halt with the input $\bsu$.
}
}

Here, we bound both $\bA$ and $\overline{\calV\calA \calV^\dagger}$ in order to treat STA and HTA simultaneously.
By this lemma, we can prove that the halting problem, a well-known undecidable problem, is not easier than STA nor HTA: 
So the latter is also undecidable.

\bigskip

We have discussed the undecidability of relaxation, where we consider whether the long-time average is close to a given value $A^*$.
In the case of thermalization, we consider whether the long-time average is close to the equilibrium value $\Tr[\calA\rho^{\rm MC}]$.
Modifying slightly the proof techniques for Lemma 1, we can obtain HTA-type undecidability of thermalization.
Here, we set $d_1=d_0+5$.

\ib{Theorem 2}{
Fix the local dimension $d\geq d_1$.
Let $A$ be an arbitrary observable on a single site and $\rho$ be an arbitrary pure state in the form of \eref{ini-form}.
We require that there exists states $\ket{\phi_2}$ and $\ket{\phi_3}$ orthogonal to each other and to $\ket{\phi_0}$, $\ket{\phi_1}$, $A\ket{\phi_0}$ and $A\ket{\phi_1}$ such that $\braket{\phi_2|A|\phi_2}>\max_{\ket{\psi}\in {\rm span}\{ \ket{\phi_0}, \ket{\phi_1}\}} \braket{\psi|A|\psi}$ and $\braket{\phi_3|A|\phi_3}<\min_{\ket{\psi}\in {\rm span}\{ \ket{\phi_0}, \ket{\phi_1}\}} \braket{\psi|A|\psi}$.

Then, for any large $M>1$ there exist proper $\ep_2$ and $\ep_1$ with $\ep_2\geq\ep_1M$ such that the HTA for $A$ and $\rho$ with $c=\Tr[\calA\rho^{\rm MC}]$, $\ep_2$ and $\ep_1$, is undecidable.
This remains valid even if the Hamiltonian $H$ (inputs of HTA) is restricted to a shift-invariant Hamiltonian $H$ with 1-body terms and 2-body nearest-neighbor interaction terms.
}

The conditions of $\ket{\phi_2}$ and $\ket{\phi_3}$ mean that the initial state $\ket{\phi_1}$ and $\ket{\phi_0}$ is not at the edge of the spectrum of $A$.
Except for such anomalous cases, almost all setups are expected to satisfy these conditions.

We clarify the dependency of quantities, which is similar to that of Theorem 1b:
Both the observable $A$ and the initial state $\rho$ are arbitrary (as far as the condition for $\ket{\phi_2}$ and $\ket{\phi_3}$ is satisfied).
Two parameters for the error, $\ep_1$ and $\ep_2$, depend on the choice of $A$ and $\rho$.
The task of HTA is to determine $\bA\in [c+\ep_1, c-\ep_1]$ or $\bA\notin [c+\ep_2, c-\ep_2]$ with $c=\Tr[\calA \rho^{\rm MC}]$ for all inputs $H$ in the form of shift-invariant and nearest-neighbor interaction, and thus to show undecidability we pick up a family of {\it bad} inputs depending on $A$, $\rho$, $\ep_1$, and $\ep_2$.

\section{Proof of Theorem 1a and Theorem 1b (undecidability of relaxation) from Lemma 1}\lb{s:pf-thm-lemma}

Below, without loss of generality, we suppose $\braket{e_1|A|e_1}=0$, which can be met by subtracting a constant factor.
We also set the target value $c$ in STA and HTA to zero.

\begin{proof}[Proof of Theorem 1a]
We fix the CONS $\{ \ket{e_i}\}$ arbitrarily.
For any given $\ep>0$ and $A$, we set the operator $A'$ as
\eq{
A'=\bcases{
\ep\ket{e_2}\bra{e_2}+A & {\rm if} \ \braket{e_2|A|e_2}\geq 0, \\
-\ep\ket{e_2}\bra{e_2}-A & {\rm if} \ \braket{e_2|A|e_2}< 0.
}
}
This choice guarantees $\braket{e_1|A'|e_1}=0$ and $\abs{\braket{e_2|A'|e_2}}\geq \ep$.
We set the system Hamiltonian $H$ as what is given in Lemma 1 with $\eta=\ep/8M$.

With these $H$ and $\eta$, we claim that the STA for the following state family for $\bsu$
\eq{
\{ \ket{\psi_{V,L}}=V\ket{e_0}\otimes (V\ket{e_1})^{\otimes L-1}|V=V_{\bsu} \} _{\bsu}
}
with $\ep_1=\eta$ and $\ep_2=(1/4-\eta)\ep$ is undecidable.
The ratio of two errors is bounded as $\ep_2/\ep_1=(1/4-\eta)\ep\cdot 8M/\ep\geq M$, where we used $\eta\leq 1/8$.
The undecidability follows from Lemma 1, which suggests that
\eq{
\bA(H, \psi_V)\geq \( \frac14 -\eta\) \braket{e_2|A'|e_2}\geq \( \frac14 -\eta\)  \ep=\ep_2
}
holds if and only if the URTM with the input $\bsu$ halts, and
\eq{
\bA(H, \psi_V)\leq \eta =\ep_1
}
holds if and only if the URTM with the input $\bsu$ does not halt.
Since the halting problem of the URTM is undecidable, this STA is also undecidable.
\end{proof}

\begin{proof}[Proof of Theorem 1b]
We set a CONS $\{ \ket{e_i}\}$ such that $\ket{e_0}$, $\ket{e_1}$, and $\ket{e_2}$ are the given states $\ket{\phi_0}$, $\ket{\phi_1}$, and $\ket{\phi_2}$, respectively, which ensures $\braket{e_2|A|e_2}\neq 0$.
Without loss of generality, we can suppose $\braket{e_2|A|e_2}>0$.
(In case of $\braket{e_2|A|e_2}<0$, we obtain an analogous conclusion by setting $A\to -A$.)
Let $H$ be the Hamiltonian whose existence is guaranteed by Lemma 1 with the above CONS, the above observable $A$, and $\eta=\braket{e_2|A|e_2}/8M$.
Using this $H$, we construct an $A$-dependent family of Hamiltonians for $\bsu$ given as
\eq{
\{ H_{V,L}=\calV^\dagger H_L \calV| \calV= V_{\bsu}^{\otimes L}\} _{\bsu} .
}

We now set $\eta=\min(\braket{e_2|A|e_2}/8M, 1/8)$ and $\ep_1=\eta$ and $\ep_2= (1/4 -\eta) \braket{e_2|A|e_2}$, which confirms $\ep_2/\ep_1\geq (1/4 -\eta) \braket{e_2|A|e_2}\cdot 8M/\braket{e_2|A|e_2}\geq M$.
By introducing a state family $\ket{\psi_{V,L}}=V_{\bsu}\ket{e_0}\otimes (V_{\bsu}\ket{e_1})^{\otimes L-1}$, we have $\bA(H_V, \psi_I)=\overline{\calV\calA \calV^\dagger}(H, \psi_V)$, and thus Lemma 1 implies that
\eq{
\bA(H_V, \psi_I)=\overline{\calV\calA \calV^\dagger}(H, \psi_V)\geq \( \frac14 -\eta\) \braket{e_2|A|e_2}=\ep_2
}
holds if and only if the URTM with the input $\bsu$ halts, and
\eq{
\bA(H_V, \psi_I)=\overline{\calV\calA \calV^\dagger}(H, \psi_V)\leq \eta =\ep_1
}
holds if and only if the URTM with the input $\bsu$ does not halt.
Since the halting problem of the URTM is undecidable, the HTA with the initial state $\ket{\psi_I}=\ket{\phi_0}\otimes \ket{\phi_1}^{\otimes L-1}$, the observable $A$, $\ep_1$ and $\ep_2$ is also undecidable.
\end{proof}

\section{Strategy for the proof of Lemma 1}\lb{s:strategy}

Most of the remainder of this \FS{paper}{Supplementary Note}, from \sref{pf1} to \sref{pf4}, is devoted to proving Lemma 1.
Before going to the details of the proof, we here sketch the strategy of the proof.

We first introduce a classical reversible Turing machine which our quantum system emulates (\sref{pf1}).
As explained in \sref{GTM}, in order to implement it by a one-dimensional quantum system we generalize conventional Turing machines such that the finite control sits in the tape.
The classical reversible Turing machine has two types of cells: M-cells and A-cells (see also \fref{rule}).
The decoding of an input code and simulation of a URTM is performed in M-cells, while A-cells are responsible for the change of $\bA$ (the long-time average of $A$) in case of halting.
Namely, we emulate the URTM in M-cells, and if and only if the URTM halts, then we start flipping the state of A-cells.
For a reason discussed later, we set the vast majority of cells to A-cells and a small fraction of cells to M-cells.

We construct a quantum system which emulates the above classical reversible Turing machine (Secs.\ref{s:pf2}-\ref{s:pf4}).
The Hamiltonian is constructed in a similar manner to the Feynman-Kitaev Hamiltonian without a clock: One-step time evolution of the classical Turing machine corresponds to an application of a quantum partial isometry as a quantum walk (\sref{pf2}).
The idea of the Feynman-Kitaev Hamiltonian is first demonstrated with a simple toy example (\sref{toy-ex}), and then we construct the quantum isometry for our classical Turing machine (\sref{q-iso-M}).

Since the dynamics of our system are a little complicated, we first introduce an analogous setting to our original one but easier to analyze, where the initial state is a single computational basis state representing the input code (\sref{pf3}).
In this analogous setting, we demonstrate how the value of $\bA$ changes depending on the halting/non-halting of the URTM.
In this case, the dynamics are fully solvable since the effective Hamiltonian for this initial state can be expressed as a tridiagonal matrix, which can be exactly diagonalized.

We then proceed to the original setting, a shift-invariant initial state (except site 1), where the initial state is a superposition of computational basis states (\sref{pf4}).
The first two subsections (\sref{ini-state} and \sref{halt1}) are devoted to describing the decoding process of the input code for the URTM from an almost uniform initial state.
With the help of the law of large numbers, we demonstrate that the input code is successfully decoded with arbitrarily high probability amplitude by choosing proper parameters.
In the subsequent two subsections, we describe how the value of $\bA$ changes in the case of halting and non-halting.
If the URTM halts, it starts flipping spins in A-cells to increase the value of $A$. 
The stable state of the corresponding quantum walk is uniformly distributed over all the states traversed by this discrete-time dynamics, so the expectation value of $A$ becomes large if the URTM halts (\sref{halt2}).
On the other hand, since the URTM is simulated on relatively small numbers of sites (M-cells), which are sparsely located in the 1D-chain, if the URTM does not halt, an overwhelming part of the system (A-cells) remains unchanged (\sref{nonhalt}).

\section{Proof of Lemma 1: (1) Classical Turing machine}\lb{s:pf1}

In this section, we construct a classical machine which is emulated by quantum many-body systems.
Our classical machine consists of three TMs, {\it TM1}, {\it TM2}, and {\it TM3}, working on two types of cells, {\it M-cells} and {\it A-cells}.
TM2 simulates a given URTM, and TM1 decodes its input from bit sequences.
TM3 is responsible for changing the value of $A$ when TM2 halts.

An M-cell has a three-layered structure:
The first layer is a working space of TM1 and TM2, and the second and third layers, which take 0 or 1, contain the information of the input code for TM2.
An A-cell has two distinct states; $a_1$ and $a_2$.

\figin{10cm}{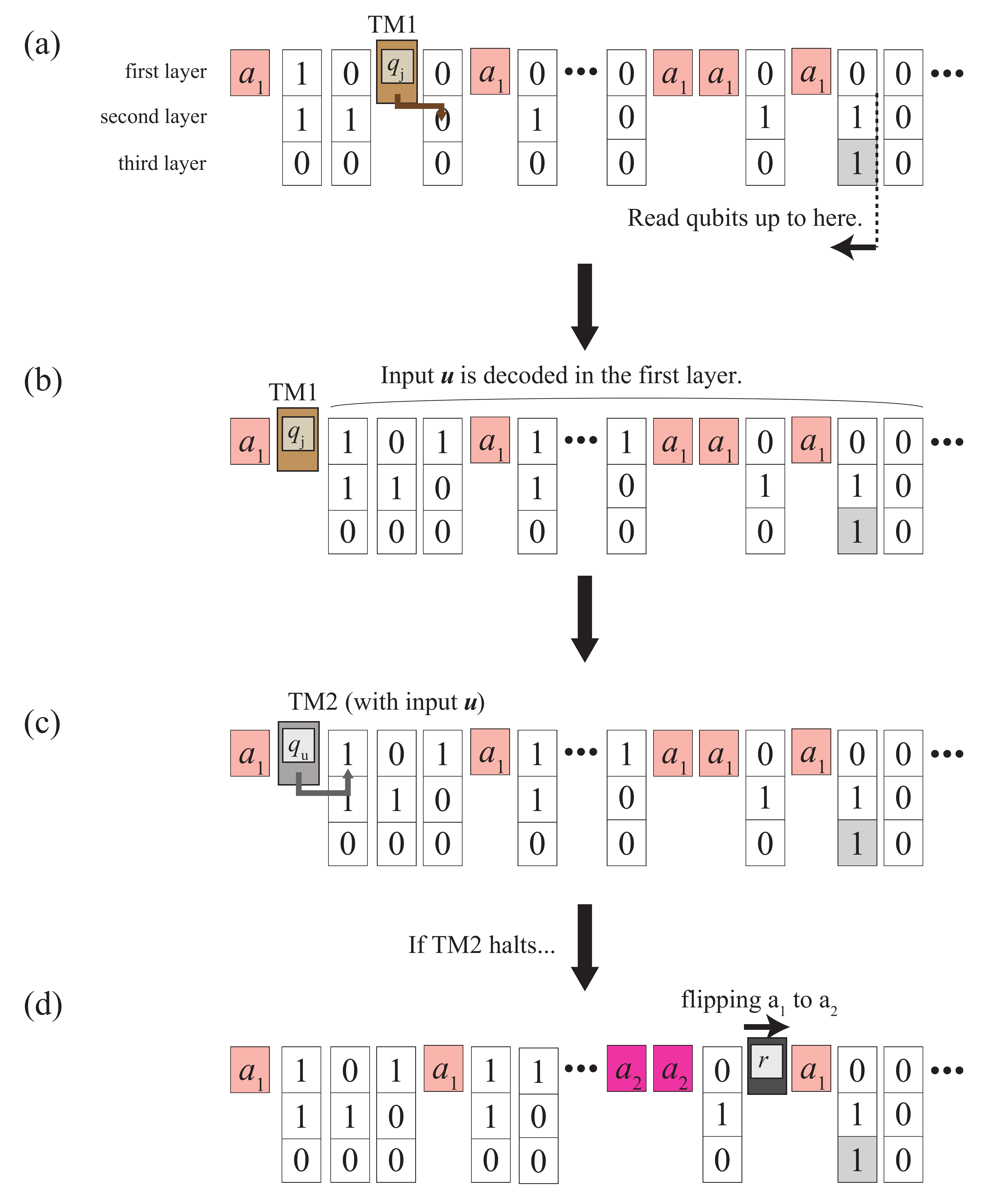}{
Schematic of the dynamics and the structure of the emulated classical machine.
Two types of cells, M-cell and A-cell, and a finite control sit in a single line.
M-cells are three-layered colored in white, and A-cells are colored in red.
In this figure, $q_j$, $q_u$, and $r$ are the internal states of TM1, TM2, and TM3, respectively.
(a) TM1 decodes the input code $\bsu$ for TM2 from the second and third layers of M-cells.
The relative frequency of 1's in the second layer in the binary expansion is set to $\bsu$, and the leftmost 1 in the third layer (colored in gray) tells how long TM1 reads cells in the second layer and how many digits it decodes.
Note that TM1 and TM2 pass through A-cells.
(b) After decoding by TM1, the input code $\bsu$ is output to the first layer of M-cells. 
(c) When TM1 stops, then a universal reversible Turing machine TM2 starts working.
If TM2 does not halt, TM2 eventually passes the periodic boundary, and then TM2 stops its move.
(d) If TM2 halts, then TM3 start flipping the state in A-cells from $a_1$ to $a_2$, which changes the value of $\bA$.
}{rule}

\subsection{Universal reversible Turing machine}

We fix a URTM characterized by the set $(Q,\Gamma, q_0, q_f, s_0, \delta)$.
Here, $Q$ is a set of states of the finite control, $\Gamma$ is a set of tape alphabet symbols, $q_0$ and $q_f$ are the initial and the final state of the finite control respectively, $s_0$ is the blank symbol, and $\delta$ is the transition function.
The set $\Gamma$ contains the symbol $\square$, which stands for the leftmost cell.
Without loss of generality, we suppose the followings:
\be{
\item We essentially use the quadruple form, that is, split each step into the two sub-steps:
The first sub-step is rewriting the internal state and the symbol in the cell, and the second sub-step is the move of the tape head.
To realize such moves of the machine, we set $Q:=Q_m\times Q_u$ with $Q_m:=\{ m_0,m_1\}$, where $q_m\in Q_m$ is used to represent the two distinct sub-steps.
It equals $m_0$ when rewriting and $m_1$ when moving the tape head.
At every step, the value of $q_m$ is flipped so that rewriting (other than the $Q_m$ part) and moving on the tape occurs alternatingly.
This is for the sake of clarity and at the same time to realize the movement of the tape head to the left by a two-body interaction.

\item We suppose that TM2 has {\it unique direction property}~\cite{Morbook}:
$Q_u$ is split into three disjoint subsets, $Q_+$, $Q_-$, and $Q_0$ (i.e., $Q_u=Q_+\cup Q_-\cup Q_0$).
The state $q_u$ is in $Q_+$ (resp. $Q_-$) if and only if the tape head has moved to the right (resp. left), and in $Q_0$ if and only if the tape head has not moved.
We promise that $q_0\in Q_m\otimes Q_+$ and $q_f\in Q_m\otimes Q_-$.
This property confirms the reversibility of this TM.

\item The tape head of the initial configuration is at the leftmost cell, and the $Q_m$ part equals $m_1$.
Hence, at the first step, it moves to the right.
}

\subsection{Generalized URTM}\lb{s:GTM}

To implement classical TMs by one-dimensional quantum many-body systems, we need some modifications to the above URTM.
First, in conventional TMs the finite control sits outside the tape, while we will set the finite control in the tape.
Second, we add two (non-universal) TMs, TM1 and TM3, which work before and after TM2 runs, respectively.

We first generalize the TM such that the finite control sits in the tape.
Any RTM (reversible Turing machine) can be simulated by the following generalized URTM, as long as the length of the tape suffices.
The generalized TM is almost the same as the URTM in the previous subsection, except for the following respects:
\bi{
\item The finite control sits between the two cells of the tape, and can read only one of the two adjacent cells in a single step.
If it reads the cell right (left) to it, it moves to the right (left) or does not move.
Each element $q$ of the extended state space ${Q}$ (we employ the same symbol for this extended space for brevity) tells which cell will interact with the finite control in this step.

\item (Unique pairing property) For the sake of reversibility, we assume the following:
Each element $q$ of the state space ${Q}$ tells the previous position of the finite control and the cell it has interacted with.
In analogy with the unique direction property of an RTM, we call this property {\it a unique pairing property}.

\item There are several configurations with no successor.
}

We next add TM1 and TM3.
Since the detailed rule of TM1 will be presented in \sref{ini-state}, we here briefly explain the dynamics of TM1 and TM3.

As explained, the second and third layers consist of 01-bit sequences.
Let $\beta$ be a real number whose binary expansion is equal to the input code $\bsu$, and set the relative frequency of 1 in the second layer to $\beta$.
We also set the third layer such that almost all cells are filled with 0, and the state 1 rarely appears.
TM1 counts the relative frequency of 1 in the second layer until we encounter the first 1 in the third layer.

If and only if TM2 halts, TM3 starts working.
TM3 moves right, and if TM3 encounters an A-cell with the state $a_1$, it flips the state to $a_2$.
If TM3 encounters an A-cell with the state $a_2$, which means that TM3 has moved around the system and goes back to the start point, then TM3 stops.

We call the composite classical machine of these TMs as generalized URTM and denote it by $\MG$.

\subsection{The states, symbols, and tape of $\MG$}

We define the set of states and symbols of $\MG$ as
\balign{
Q\cup Q_r&=(Q_m\times Q_u)\cup Q_r, \\
\overline{\Gamma} \times \Gamma_{\rm in}\times \Gamma_A&=(\Gamma\cup \{ s_1\}) \times (\{ 0,1\}^{\otimes 2}\cup \{ \psi_0\} )\times \Gamma_A,
}
where we set
\balign{
Q_m&:=\{ m_0,m_1\} , \\
Q_r&:=\{ r\}, \\
\overline{\Gamma}&:=\Gamma \cup \{ s_1\} , \\
\Gamma_A&:=\{ a_1,a_2\}.
}
Here, $Q_m\times Q_u$ represents the states of the finite control of TM1 and TM2, and $Q_r=\{ r\}$ is the sign for TM3, which flips the states in A-cells.
The alphabets $\Gamma$, $\Gamma_{\rm in}$ and $\Gamma_A$ appear in the first layer of M-cells, the second and third layers of M-cells, and A-cells, respectively.
The symbol $s_1$ and $\psi_0$ are signs that this cell is an A-cell.
Note that $\Gamma$ contains the blank symbol $s_0$.
We compile these symbols in \fref{state}.

\figin{14cm}{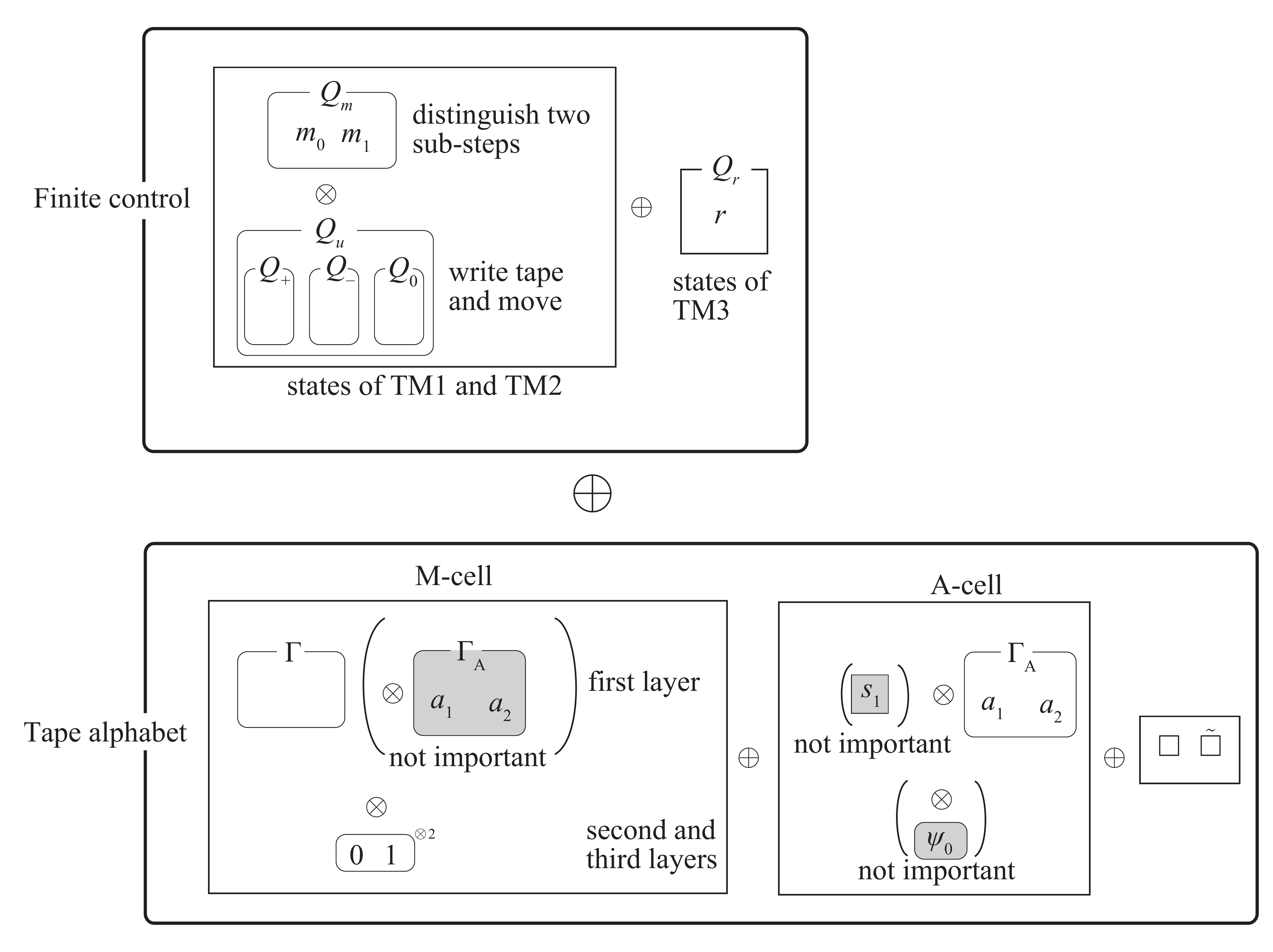}{
Summary of possible states of a single site in the quantum system.
The symbols in gray areas are dummy states which are not important.
}{state}

Out of the $L-1$ cells, $\lfloor \alpha (L-1)\rfloor$ cells are filled with symbols from $\Gamma\times \Gamma_A$, which serve as M-cells, and $\lceil (1-\alpha)(L-1)\rceil$ cells are filled with $\{ s_1\} \times \Gamma_A$, which serve as A-cells.
M-cells simulate TM1 and TM2, and A-cells inflate the value of the observable $A$ in case of halting.
The rate $\alpha$ will be taken sufficiently small to observe the gap between the halting case and the non-halting case (i.e., almost all cells are set as A-cells, though the number of M-cells is also sufficiently large in $L\to \infty$ limit).

Initially, $\MG$'s state is set to $(m_0, q_0)$, and the $\Gamma_A$-part of the cell is set to $a_1$. 
Cells in the left of the finite control are set to the blank cell $s_0$.
Below, for brevity we leave off the symbols $\lfloor \cdot \rfloor$ and $\lceil \cdot \rceil$ and also drop $-1$, as $L$ is large enough (i.e., we write $\alpha L$ and $(1-\alpha)L$ instead of $\lfloor \alpha (L-1)\rfloor$ and $\lceil (1-\alpha)(L-1)\rceil$).

\subsection{The move of $\MG$: the first step}

When $\MG$ is in the initial state $(m_0, q_0)$, this RTM not only mimics the move of TM1 but also prepares the cell with $\square$.
The finite control reads the site on its right and rewrites the cell as $s_0\to \square$ and $s_1\to \square'$ (There is no other possibility in legal configurations).
We employ two different symbols $\square$ and $\square'$ only for satisfying the reversibility, and $\square$ and $\square'$ plays completely the same role: telling the left/right end of the tape.
Both TM1 and TM2 stop if they hit these cells, while TM3 just passes this cell transparently.
Therefore, in the remainder of this \FS{paper}{Supplementary Note}, we do not distinguish $\square'$ from $\square$.

At the first step, the state and the position of the finite control will be changed in the same manner as the TM1's first step.
Hence, first, it evolves to $(m_1,q)$, and then it moves to the right and evolves to $(m_0,q)$.

\subsection{The move of $\MG$: simulating TM1 and TM2}

While the state of the finite control $q$ is in $Q$, $\MG$ simulates TM1 and TM2 in M-cells, and it virtually ignores A-cells (see \fref{rule}.(a)-(c)).
\bi{
\item If $q_m=m_0$, it reads the cell in its right.
If it is an M-cell, the machine updates its state of $Q_u$ part, $q_u$ say, and the symbol in the cell according to the transition function $\delta$.
At the same time, $Q_m$ part of the finite control, $q_m$ say, evolves to $m_1$.
If it is an A-cell, $q_u$ does not change and $q_m$ simply evolves to $m_1$.
\item If $q_m=m_1$ and $q_u\in Q_+$, the finite control is swapped with the cell in its right, and $q_m$ evolves to $m_0$.
\item If $q_m=m_1$ and $q_u\in Q_-$, the finite control is swapped with the cell in its left, and $q_m$ evolves to $m_0$.
\item If $q_m=m_1$ and $q_u\in Q_0$, the finite control does not move, and $q_m$ simply evolves to $m_0$.
}
If $q_m=m_1$, the previous partner of the finite control is the cell on its right, and its previous position is the same as the present position.
If $q_m=m_0$ and $q_u\in Q_+$ (resp. $q_u\in Q_-$), the previous partner of the finite control is the cell on the left (resp. right) of it, and the previous position is at the left (resp. right) of the previous partner.

Eventually, the machine may run out of the tape:
In case of the periodic boundary condition, the finite control reads the symbol $\square$ with $q_u\in Q_+\backslash \{ q_0\}$.
Since these states are illegal configurations, no successor is defined to them.

If the machine runs into the halting state $q_f$, then the finite control evolves into the state $r\in Q_r$.
The dynamics after halting are given in the next subsection.

\subsection{The move of $\MG$: simulating TM3}

If TM2 halts and the state of the finite control becomes in $q\in Q_r$ (i.e., $q=r$), then TM3 starts flipping the state in A-cells from $a_1$ to $a_2$ (\fref{rule}.(d)).
If the cell on the right of the finite control is an A-cell with its state $a_1$, then the head flips the state to $a_2$ and moves right.
If the cell on the right of the finite control is an M-cell, then the finite control just moves right.
If the cell on the right of the finite control is an A-cell with its state $a_2$, which implies that all of the A-cells have already been flipped to $a_2$ and the finite control has gone around the periodic system, then TM3 stops.

\section{Proof of Lemma 1: (2) Quantum partial isometry corresponding to $\MG$}\lb{s:pf2}

\subsection{Toy example of Feynman-Kitaev type Hamiltonian}\lb{s:toy-ex}

Before constructing the quantum partial isometry for $\MG$, for readers who are not familiar with the emulation of classical machines by quantum systems, we here demonstrate a toy example of a quantum partial isometry emulating a very simple classical machine.
If a reader is familiar with this topic, one can skip this subsection and start reading the subsection entitled {\it Construction of quantum partial isometry for $\MG$}.

\bigskip

\figin{12cm}{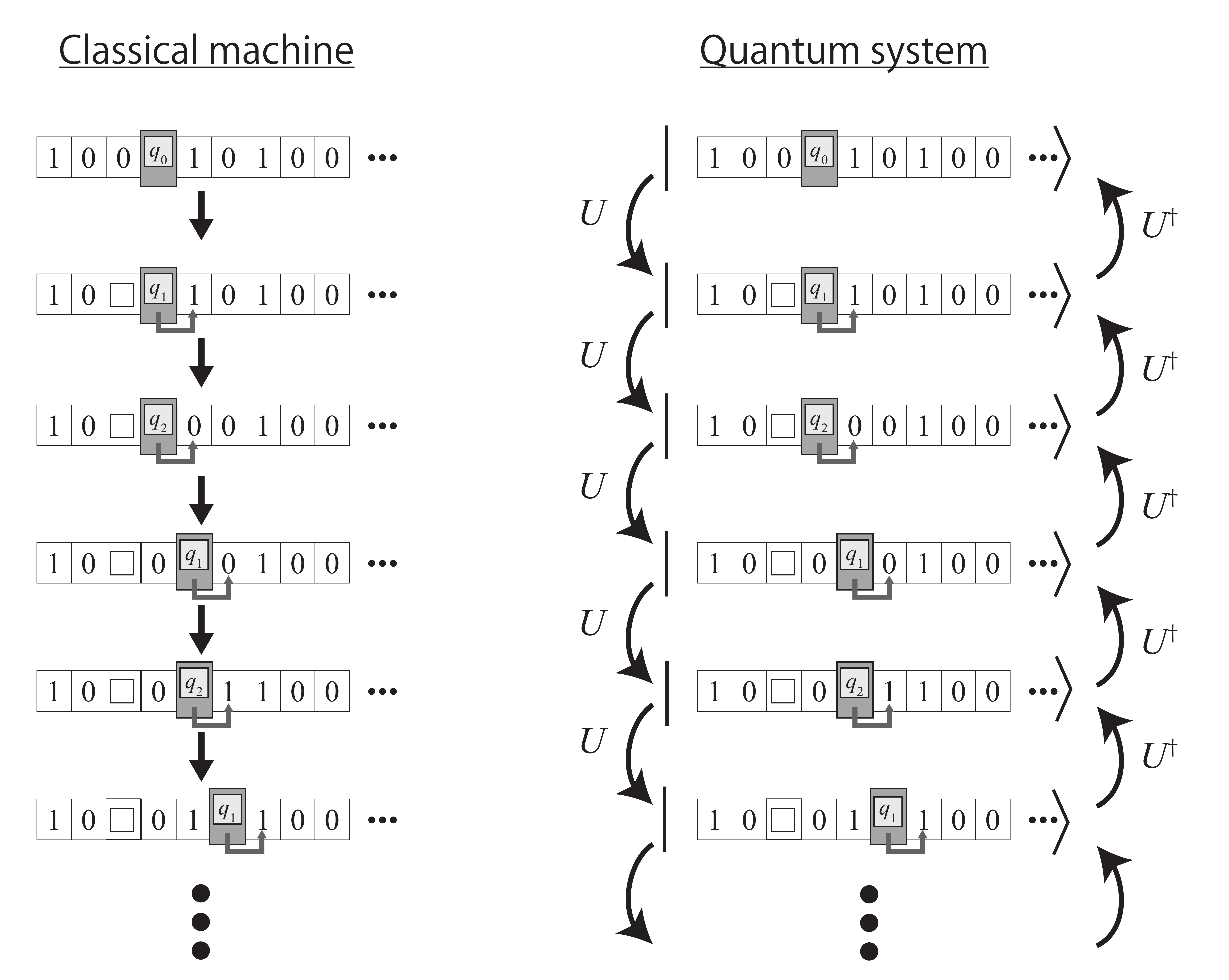}{
The dynamics of the toy example of a classical machine introduced in \sref{toy-ex}, and the corresponding quantum partial isometry $U$.
}{qiso}

We consider a simple classical machine, flipping bits with moving from left to right on a single cell line.
The finite control also settles in this line.
The finite control takes three possible states, $q_0$, $q_1$ and $q_2$, and the transition rule of this machine is the following (see also \fref{qiso}):
\be{
\item If the state of the finite control is $q_0$, then the state of the finite control evolves to $q_1$.
At the same time, the cell on the left of the finite control is flipped to $\square$ (if the cell is 0) or $\tilde{\square}$ (if the cell is 1).
\item If the state of the finite control is $q_1$ and the cell on the right of the finite control is neither $\square$ nor $\tilde{\square}$, then the machine flips the bit (i.e., $0\to 1$ and $1\to 0$) in the cell on the right of the finite control.
At the same time, the state of the finite control evolves to $q_2$.
\item If the state of the finite control is $q_2$, then the finite control moves to the right.
At the same time, the state of the finite control evolves to $q_1$.
\item If the state of the finite control is $q_1$ and the cell on the right of the finite control is $\square$ or $\tilde{\square}$, then the machine stops.
}

We emulate the above classical machine by a one-dimensional quantum system with nearest-neighbor interaction.
The local Hilbert space is spanned by the following seven states:
\eq{
\{ \ket{0}, \ket{1}, \ket{q_0}, \ket{q_1}, \ket{q_2}, \ket{\square}, \ket{\tilde{\square}}\}. \nt
}
The local quantum isometry on the sites $i$ and $i+1$ is given as
\eq{
U_{i,i+1}=U^1_{i,i+1}+U^2_{i,i+1}+U^3_{i,i+1}
}
with
\balign{
U^1=&\ket{\square q_1}\bra{0q_0}+\ket{\tilde{\square}q_1}\bra{1 q_0}, \\
U^2=&\ket{q_2 1}\bra{q_1 0}+\ket{q_2 0}\bra{q_1 1}, \\
U^3=&\ket{1 q_1}\bra{q_2 1}+\ket{0 q_1}\bra{q_2 0},
}
which correspond to the transition rules 1,2,3, respectively.
The quantum isometry of the total system is given by
\eq{
U=\sum_i U_{i,i+1}.
}

The initial state is restricted to the form that only a single site takes $\ket{q_0}$ and other sites are in the local Hilbert subspace spanned by $\{ \ket{0}, \ket{1}\}$, which ensures the emulation of legal states of the classical machine.
By denoting by $\ket{\psi_n}$ the state representing the state of the classical machine at the $n$-th step, then 
\eq{
U\ket{\psi_n}=\ket{\psi_{n+1}}
}
is satisfied, which means that the quantum isometry $U$ induces one step evolution of the quantum state.

\subsection{Construction of quantum partial isometry for $\MG$}\lb{s:q-iso-M}

Now we consider our original Turing machine.
By defining
\eq{
X:=Q\cup Q_r\cup \{ \overline{\Gamma}\times \Gamma_A\} ,
}
then the sequence of the elements $\bsx=(x_1,x_2,\ldots , x_L)$ ($x_i\in X$) represents an instantaneous configuration of $\MG$, where only one of $x_i$'s is a member of $Q\cup Q_r$.
We here dropped the second and third layers of M-cells (i.e., $\Gamma_{\rm in}$ from $\{ \overline{\Gamma}\times \Gamma_A\}$) for brevity, which are explicitly treated in \sref{ini-state}.
The last condition on $\bsx$ ensures a single finite control and a single tape head in the system.

We denote by $\calH^X$ the Hilbert space spanned by $\{ \ket{x}|x\in X\}$, and define $\calH^Q$, $\calH^{Q_m}$, $\calH^{Q_u}$, $\calH^{Q_r}$, $\calH^{\overline{\Gamma}}$ and $\calH^{\Gamma_A}$ in a similar manner, which satisfy
\balign{
\calH^X&=\calH^Q\oplus \calH^{Q_r}\oplus (\calH^{\overline{\Gamma}}\otimes \calH^{\Gamma_A}), \\
\calH^Q&= \calH^{Q_m}\otimes\calH^{Q_u}.
}
A quantum state $\ket{\bsx}\in (\calH^X)^{\otimes L}$ represents $\bsx$, a configuration of $\MG$.
Since the states in $\calH^X$ are only the label (except for $\Gamma_A$), we assign the CONS given in Lemma 1 to the CONS $\{ \ket{x}| x\in X\}$ as
\balign{
\ket{m_0, q_0}&=\ket{e_0}, \\
\ket{s_1, a_i}&=\ket{e_i}, \ \  (i=1,2).
}
With this assignment, the value of $A$ indeed varies between $i=1$ and $i=2$ with the state $s_1$:
\balign{
\braket{s_1,a_1|A|s_1,a_1}&=\braket{e_1|A|e_1}=0, \\
\braket{s_1,a_2|A|s_1,a_2}&=\braket{e_2|A|e_2}>0.
}

Each step of the move (time evolution) of $\MG$ can be described by a partial isometry acting on at most two sites simultaneously.
Thus, the sum of the partial isometries acting on at most two sites properly emulates the time evolution of $\MG$.
Suppose that in $\MG$ the tape head is at the $i$-th cell, or equivalently $x_i\in Q\cup Q_r$.
We set the partial isometry $U_i$ acting on the $i-1$, $i$, and $i+1$-th sites as
\eqa{
U_i=U_i^0+U_i^{1+}+U_i^{1-}+U_i^{10}+U_i^{r},
}{def-Ui}
where each summand acts on at most two sites.
If the state of the finite control is in $(m_0, q)$, the move of the machine is emulated\fn{
This is represented by an isometry from $\bbC (\ket{m_0}\otimes \calH^{Q_u})_i \otimes (\calH^{\overline{\Gamma}}\otimes \calH^{\Gamma_A})_{i+1}$ to the same Hilbert space.
}
by $U_i^0$.
If the $i$-th and $i+1$-th cells in $\MG$ evolve from $(x_i, x_{i+1})$ to $(x'_i, x'_{i+1})$, the corresponding isometry on the sites $i$ and $i+1$ is given by $\ket{x'_i, x'_{i+1}}\bra{x_i, x_{i+1}}$.
If the state of the finite control is in $(m_1,q)$, the move is emulated by one of $U_i^{1+}$, $U_i^{1-}$, or $U_i^{10}$, depending on whether $q\in Q_+$, $Q_-$, or $Q_0$.
They swap the sites and change $m_1$ to $m_0$.
$U_i^r$ mimics the move of TM3 if the state is $r\in Q_r$ and the finite control is at $i$-th site.

We note that if the $i$-th site does not correspond to the finite control (i.e., $x_i\notin Q\cup Q_r$), then $U_i\ket{\bsx}=0$, which ensures the fact that only the vicinity of the finite control can change.
Owing to this, the isometry on $(\calH^X)^{\otimes L}$ given by
\eq{
U=\sum_i U_i 
}
properly emulates the dynamics of the generalized RTM $\MG$.

\section{Proof of Lemma 1: (3) Evaluating $\bA$ for computational basis state}\lb{s:pf3}

Since the computation of $\bA$ from the almost uniform initial state $\ket{\phi_0}\otimes \ket{\phi_1}^{\otimes L-1}$ is a little complicated, in this section we first consider an easier setting with a computational basis initial state.
With this initial state, 
We treat almost uniform initial states in the next section.

\subsection{General expression of long-time average}

We first derive a general expression of the long-time average $\bA$.
Let $H$ be a Hamiltonian of the system, and $E_i$ and $\ket{E_i}$ be eigenenergy and corresponding energy eigenstate.
We expand the initial state $\ket{\psi}$ with the energy eigenbasis as $\ket{\psi}=\sum_i c_i \ket{E_i}$.
Then, the long-time average of an observable $A$ is calculated as
\balign{
\bA&=\lim_{T\to \infty}\frac1T \int_0^T dt \braket{\psi(t)|A|\psi(t)} \nt \\
&=\lim_{T\to \infty}\frac1T \int_0^T dt \sum_{i,j} e^{-i(E_j-E_i)t}c_i^*c_j\braket{E_i|A|E_j} \nt \\
&=\sum_{i,j} \chi(E_j=E_i) c_i^*c_j \braket{E_i|A|E_j}, \lb{time-ave}
}
where $\chi(E_j=E_i)$ takes 1 if $E_j=E_i$ and takes zero otherwise.
We set the Planck constant to unity.
In the third line, we used the fact that $\lim_{T\to \infty}\frac1T \int_0^T dt e^{-i(E_j-E_i)t}$ converges to zero if $E_j\neq E_i$ and equal to 1 if $E_j=E_i$.

In particular, if the Hamiltonian $H$ has no degeneracy, we have a simple expression $\bA=\sum_i \abs{c_i}^2 \braket{E_i|A|E_i}$.
In contract, if there exist degeneracy ($E_i=E_j$ with $i\neq j$), we need to handle off-diagonal elements $\braket{E_1|A|E_j}$.

\subsection{Hamiltonian emulating $\MG$ and effective Hamiltonian}

We now construct the Hamiltonian of the quantum system emulating $\MG$.
We define the Hamiltonian as
\eq{
H=\sum_i (U_i+U_i^\dagger),
}
where $U_i$ is introduced in \eref{def-Ui}.

Let $\bsx^j$ be the configuration of $\MG$ at the $j$-th step, and $J$ be the number of time steps until the machine stops.
Then, its corresponding quantum state $\ket{\bsx^j}$ satisfies
\eq{
\ket{\bsx^j}=U^{j-1}\ket{\bsx^1},
}
and our Hamiltonian restricted to the Hilbert space spanned by $\{ \bsx^i\}_{i=1}^J$ can be expressed as
\eq{
\Heff:=\sum_{j=1}^{J-1} \ket{\bsx^{j+1}}\bra{\bsx^j}+{\rm h.c.}.
}
Note that $\bsx^J$ does not have a successor, and hence the evolution of $\MG$ stops at this point.

Since $\Heff$ is a tridiagonal matrix with the basis $\{ \ket{\bsx^i}\}_{i=1}^J$, we can fully solve its eigenenergies and eigenstates.
The $k$-th eigenenergy is written as
\eqa{
E_k=2\cos \( \frac{k\pi}{J+1}\)
}{exact-E}
($k=1,2,\ldots , J$) with the corresponding energy eigenstate
\eqa{
\ket{E_k}=\sqrt{\frac{2}{J+1}} \sum_{j=1}^{J}\sin \( \frac{kj\pi}{J+1}\) \ket{\bsx^j}.
}{exact-psi}
Note that $0<\frac{k\pi}{J+1}<\pi$ guarantees the absence of degeneracy in this Hamiltonian.

\subsection{Computing long-time average of $\calA$}

The initial state $\ket{\bsx^1}$ is expanded by the energy eigenstates as
\eq{
\ket{\bsx^1}=\sqrt{\frac{2}{J+1}} \sum_{k=1}^{J}\sin \( \frac{k\pi}{J+1}\) \ket{E_k},
}
and thus  $\bA$ is calculated as
\balign{
\bA&=\frac{2}{J+1} \sum_{k=1}^{J}\sin^2 \( \frac{k\pi}{J+1}\) \braket{E_k|\calA_L|E_k} \nt \\
&=\( \frac{2}{J+1}\) ^2 \sum_{j,j'=1}^J \[  \sum_{k=1}^{J}\sin^2 \( \frac{k\pi}{J+1}\) \sin \frac{j'k\pi}{J+1} \sin \frac{jk\pi}{J+1}\] \braket{\bsx^j|\calA_L|\bsx^{j'}} \nt \\
&=\frac{3}{2(J+1)}\( \braket{\bsx^1|\calA_L|\bsx^1}+\braket{\bsx^J|\calA_L|\bsx^J}\) +\frac{1}{J+1}\sum_{j=2}^{J-1}\braket{\bsx^j|\calA_L|\bsx^j}-\frac{1}{2(J+1)}\sum_{\substack{1\leq j, j'\leq J \\ j=j'\pm 2}}\braket{\bsx^j|\calA_L|\bsx^{j'}}. \lb{bA-single-last}
}
Here, in the first line we used \eref{time-ave}, and in the last line we used the following relation:
\eq{
\sum_{k=1}^{J}\sin^2 \( \frac{k\pi}{J+1}\) \sin \frac{j'k\pi}{J+1} \sin \frac{jk\pi}{J+1}=
\bcases{
\frac14 (J+1) &j=j'\neq 1 \ {\rm and} \ \neq J, \\
\frac38 (J+1) &j=j'=1 \ {\rm or} \ J, \\
-\frac18 J(J+1) & j=j'\pm 2, \\
0 &j\neq j' \ {\rm and} \ j\neq j'\pm 2.

}
}
Now, we shall show that the last term $\braket{\bsx^j|\calA_L|\bsx^{j'}}$ is sufficiently small.
Since $\bsx^j\neq \bsx^{j'}$ and $\calA_L$ is a sum of one-body observables; $\calA_L=\frac1{L}\sum_i A_i$, we find that $\braket{\bsx^j|\calA_L|\bsx^{j'}}$ can take a nonzero value only if $\bsx^j$ and $\bsx^{j'}$ differs only in a single site.
By denoting this site by $i^*$, we bound the last term $\braket{\bsx^j|\calA_L|\bsx^{j'}}$ as
\balign{
\abs{\braket{\bsx^j|\calA_L|\bsx^{j'}}}=\frac{1}{L}\abs{\sum_{i=1}^{L}\braket{\bsx^j|A_i|\bsx^{j'}}}=\frac{1}{L}\abs{\braket{\bsx^j|A_{i^*}|\bsx^{j'}}}\leq \frac{1}{L}\|A\| \lb{nonuni-off}.
}
Thus, the last term of \eref{bA-single-last} is bounded above as
\eq{
\abs{\frac{1}{2(J+1)}\sum_{\substack{1\leq j, j'\leq J \\ j=j'\pm 2}}\braket{\bsx^j|\calA_L|\bsx^{j'}}}\leq \abs{\frac{1}{2(J+1)}\sum_{\substack{1\leq j, j'\leq J \\ j=j'\pm 2}}\frac{1}{L}\|A\| }=\frac{J-2}{J+1}\frac{1}{L}\|A\| ,
}
whose right-hand side vanishes in the $L\to \infty$ limit.
Therefore, in the following, we dropped this term for brevity.

We finally evaluate the first two terms of \eref{bA-single-last}.
Let $N_A(\bsx^j)$ be the number of A-cells in $\bsx^j$ filled with $a_2$.
Then, we have 
\eqa{
\abs{\braket{\bsx^j|\calA_L|\bsx^j}-\frac{N_A(\bsx^j)}{L}\braket{e_2|A|e_2}}\leq \alpha \|A\|,
}{nonuni-correction}
where the difference between $\braket{\bsx^j|\calA_L|\bsx^j}$ and $\frac{N_A(\bsx^j)}{L}\braket{e_2|A|e_2}$ comes from the presence of M-cells.
In the case of halting, by taking $L$ sufficiently large, we can make $\MG$ halt before $J/2$ steps.
In this condition, $\frac{N_A(\bsx^j)}{L}\geq 2(j-\frac J2)$ is satisfied for $j\geq \frac J2$, which indicates
\eqa{
\bA=\frac{3}{2(J+1)}\( \braket{\bsx^1|\calA_L|\bsx^1}+\braket{\bsx^J|\calA_L|\bsx^J}\) +\frac{1}{J+1}\sum_{j=2}^{J-1}\braket{\bsx^j|\calA_L|\bsx^j} \geq \frac14 \braket{e_2|A|e_2}-\alpha \|A\| .
}{time-ave-nonuni-fin}
By taking $\alpha$ sufficiently small, we arrive at the relation $\bA\geq \( \frac14-\eta\) \braket{e_2|A|e_2}$ in case of halting.

In contrast, in case of non-halting, since $N_A(\bsx^j)=0$ for any $j$, we have
\eq{
\bA=\frac{3}{2(J+1)}\( \braket{\bsx^1|\calA_L|\bsx^1}+\braket{\bsx^J|\calA_L|\bsx^J}\) +\frac{1}{J+1}\sum_{j=2}^{J-1}\braket{\bsx^j|\calA_L|\bsx^j} \leq \alpha \|A\|.
}
By taking $\alpha$ sufficiently small, we arrive at the relation $\bA\leq \eta$ in case of non-halting.

\section{Proof of Lemma 1: (4) Evaluating $\bA$ for superposition of computational basis states}\lb{s:pf4}

\subsection{Setting of the initial state and decoding of the input}\lb{s:ini-state}

We now consider our original setting where the initial state is shift-invariant except the first site, which takes the form of
\eq{
\ket{\psi_{V,L}}=(V\ket{e_0})\otimes (V\ket{e_1})^{\otimes L-1}=\ket{e_0}\otimes (V\ket{e_1})^{\otimes L-1}.
}
Here, $\ket{e_0}$ represents the state corresponding to the initial state of the finite control, and others represent cells.

As announced in \sref{q-iso-M}, we first elongate the local Hilbert subspace to $\calH^{\overline{\Gamma}}\otimes \calH^{\Gamma_A}\otimes \Hin$ in order to treat the alphabets in $\Gamma_{\rm in}$ in $\MG$, which represent the second and third layers of M-cells.
The Hilbert space $\Hin$ is a 5-dimensional space which is a sum of a 2 qubit space and a single state space $\{ \ket{\psi_0}\}$:
\eq{
\Hin ={\rm span}\{ \{\ket{0}, \ket{1}\} ^{\otimes 2} \oplus  \ket{\psi_0}\} .
}
The input code for TM2 is encoded into $\ket{\psiin}\in\{ \ket{0}, \ket{1}\} ^{\otimes 2} \subset \Hin$, which sits in all M-cells.
If the cell is an A-cell, this part is blank (constant independent of the input) denoted by $\ket{\psi_0}\in \Hin$.

In the previous section, we set $\alpha L$ cells to M-cells and $(1-\alpha)L$ cells to A-cells deterministically.
In this section, instead of this, we set $V\ket{e_1}$ as a superposition of these two types of cells.
We first assign $\ket{e_1}$ and $\ket{e_2}$ as
\eqa{
\ket{s_1, a_k}\ket{\psi_0}=\ket{e_k}\in \calH^{\overline{\Gamma}}\otimes \calH^{\Gamma_A}\otimes \Hin , \ \ (k=1,2),
}{ek-def}
and then apply an operator $V$ which slightly rotates $\ket{e_k}$ ($k=1,2$) as
\eqa{
V\ket{e_k}=\sqrt{\alpha}\ket{s_0, a_k}\ket{\psiin}+\sqrt{1-\alpha}\ket{e_k}.
}{V-def}
Here, two states, $\ket{s_0, a_1}\ket{\psiin}$ and $\ket{e_1}=\ket{s_1,a_1}\ket{\psi_0}$, correspond to the initial states of M-cells and A-cells, respectively (see also \fref{state}).
The symbol $s_0$ serves as a blank cell of M-cells in the first layer, and $\ket{\psiin}\in \Hin$ stores the input $\bsu$ for TM2 in the second and third layers of M-cells.
In the state $\ket{s_0, a_k}\ket{\psiin}$, the symbol $a_k$ plays no role.
In the state $\ket{s_1,a_k}\ket{\psi_0}$, the symbol $s_1$ is a sign to be A-cells, and $\ket{\psi_0}$ plays no role.
The symbols $a_1$ and $a_2$ distinguish two states of A-cells, which change the value of $A$.
We set the operator $V$ as an identity operator on the orthogonal component of $\calH^{\overline{\Gamma}}\otimes \calH^{\Gamma_A}\otimes \Hin$.
In particular, $V$ stabilizes $\ket{e_0}$ (i.e., $V\ket{e_0}=\ket{e_0}$).

The state $\ket{\psiin}$  takes the form of
\eq{
\ket{\psiin}=(\sqrt{\beta}\ket{1}+\sqrt{1-\beta}\ket{0})\otimes (\sqrt{\gamma}\ket{1}+\sqrt{1-\gamma}\ket{0}),
}
where the binary expansion of $\beta$ is set to be equal to the input code $\bsu$ in the form of a binary bit string.
The amount of $\beta$ is guessed by the relative frequency of 1's in the second layer (see also \fref{rule}.(a)).
The second layer is a superposition of computational basis states, and thus TM1 runs in each computational basis state as a quantum superposition.
Consider $m$ copies of $\sqrt{\beta}\ket{1}+\sqrt{1-\beta}\ket{0}$, which is expanded as
\eq{
(\sqrt{\beta}\ket{1}+\sqrt{1-\beta}\ket{0})^{\otimes m}=\sum_{\bsw \in \{ 0,1\} ^{\otimes m}} \sqrt{\beta}^{N_1(\bsw)}\sqrt{1-\beta}^{m-N_1(\bsw)}\ket{\bsw}.
}
Here, $N_1(\bsw)$ is the number of 1's in the binary sequence $\bsw$.
The probability amplitude for a computational basis state $\ket{\bsw}$ is $\abs{c_{\bsw}}^2=\beta^{N_1(\bsw)}(1-\beta)^{m-N_1(\bsw)}$.
Due to the law of large numbers, the probability amplitude for states where the relative frequency of 1's is close to $\beta$ converges to 1 in the large $m$ limit:
\eq{
\lim_{m\to \infty}\sum_{\bsw: \frac{N_1(\bsw)}{m}\simeq \beta}\abs{c_{\bsw}}^2=1,
}
where the precise meaning of the symbol $\frac{N_1(\bsw)}{m}\simeq \beta$ is clarified shortly (in \eref{outputlength}).
Hence, if $m$ is sufficiently large compared to the length of the input code, TM1 guesses $\beta$ correctly from the frequency of 1's.

The length of qubit $m$ is determined by another bit sequence $\sqrt{\gamma}\ket{1}+\sqrt{1-\gamma}\ket{0}$ in the third layer.
For any given accuracy $0<\xi<1$, the information of $m$ is encoded to $\gamma$ as satisfying
\eqa{
(1-\gamma)^m \geq 1-\xi.
}{gamma-cond}
As $\gamma$ set to extremely close to 0, almost all qubits are $\ket{0}$ in this sequence, and the state $\ket{1}$ rarely appears.
In particular, $\ket{1}$ appears only after $m$-th digit with probability larger than $1-\xi$.
Owing to this, if $\ket{1}$ appears at the $m'$-th digit for the first time, this is taken as the sign of $m\leq m'$.
Based on the observed value $m'$, the length of the output by TM1 (i.e., the presumed length of the digit of $\beta$) is determined as $n'=\lceil\frac14 \log_2 m'\rceil$, which ensures
\eqa{
\lim_{m'\to \infty} {\rm Prob} \[ \abs{\frac{N_1(\bsw)}{m'}-\beta}<\frac{1}{2^{n'+1}} \] =1.
}{outputlength}
With this choice of the output length $n'$, guessing $m$ larger than the true value does not affect the correctness of the estimation of $\beta$.

\subsection{Case when TM2 halts 1: decoding and expression of states}\lb{s:halt1}

Suppose that TM2 halts on the encoded input.
We claim that in this case, with overwhelming probability amplitude the dynamics by ${U}$ before TM2 halts are described by the configuration of the first $L_0$ of M-cells, where $L_0$ is sufficiently large in the following sense:
\bi{
\item It is not smaller than the space used by TM1 and TM2.
\item It should be large enough to encounter at least a single $\ket{1}$ in the third layer in the first $L_0$ of M-cells with high probability.
In other words, for a given $0<\xi, \xi'<1$ we set $\gamma$ and $L_0$ such that
\balign{
(1-\gamma)^m &\geq 1-\xi, \nt \\
(1-\gamma)^{L_0} &\leq 1-\xi',
}
where the first line is the same as \eref{gamma-cond}.
By setting $\xi, \xi'\ll 1$, the first $\ket{1}$ in the second layer appears between the first $m$ of M-cells and the first $L_0$ of M-cells with high probability.
}
We emphasize that $L_0$ is independent of the system size $L$.

Let $C$ be the cluster of the first $\abs{C}$ sites with
\eq{
\abs{C}=\frac{L_0}{\alpha}+o(L_0) .
}
Here, the $o(L_0)$ term is chosen so that $C$ contains at least $L_0$ of M-cells with high probability.
The cells outside of the cluster $C$ may interact with the finite control only when $q\in Q_r$, which means that only $\GA$-parts may touch.
This fact motivates the following representation of the initial state:
Let $\bsy$ be the first $\abs{C}$ components of $\bsx$, that is, the configuration of sites in the cluster $C$, and denote by $Y$ the set of all possible $\bsy$'s.
Then, the initial state of the total system can be expressed as
\eq{
\ket{\psi_{V,L}}=\sum_{\bsy\in Y}c_{\bsy} \ket{\bsy}\otimes (V\ket{e_1})^{\otimes L-\abs{C}}.
}

We call $\bsy$ ``good" initial configurations if (i) TM1 correctly decode the input code $\bsu$ for TM2, and (ii) in the dynamics induced by the isometry ${U}$ the finite control does not come out from the cluster $C$ before TM2 reaches the halting state.
We write as $Y^*$ the set of good initial configurations in the above sense, and define the following unnormalized state:
\eqa{
\ket{\psi_{V,L}^*}:=\sum_{\bsy\in Y^*}c_{\bsy} \ket{\bsy}\otimes (V\ket{e_1})^{\otimes L-\abs{C}}.
}{psip}
By construction, states outside $Y^*$ have negligibly small probability weight, and thus for any $\delta>0$ we can prepare $\abs{\braket{\psi_{V,L}|\psi_{V,L}^*}}>1-\delta$.
The dynamics starting from $\ket{\psi_{V,L}^*}$ and $\ket{\psi_{V,L}}$ are arbitrarily close to each other because the trace norm between the states $\ket{\psi_{V,L}^*(t)}=e^{-i\tH t}\ket{\psi_{V,L}^*}$ and $\ket{\psi_{V,L}(t)}=e^{-i\tH t}\ket{\psi_{V,L}}$ is bounded as
\balign{
\| \ket{\psi_{V,L}^*(t)}\bra{\psi_{V,L}^*(t)}-\ket{\psi_{V,L}(t)}\bra{\psi_{V,L}(t)}\|_1=&\| \ket{\psi_{V,L}^*}\bra{\psi_{V,L}^*}-\ket{\psi_{V,L}}\bra{\psi_{V,L}}\|_1 \nt \\
\leq& 2\sqrt{1-\abs{\braket{\psi_{V,L}|\psi_{V,L}^*}}^2} \nt \\
\leq& 2\sqrt{2\delta}.
}
Since $\delta$ can be arbitrarily small, in the following we regard $\ket{\psi^*}$ as the initial state itself.

For each good initial configuration $\bsy$, we define the $j$-th state as
\eq{
\ket{j,\bsy}:={U}^{j-1}\ket{\bsy}\otimes (V\ket{e_1})^{\otimes L-\abs{C}}.
}
Recalling that ${U}$ updates the $\Gamma$-part of the M-cell in simulating TM1 and TM2, and $\GA$ part after the halting, we find that ${U}$ either changes $\bsy$ into another configuration $\bsy'$, or interchanges $V\ket{e_k}$'s ($k=1,2$).
The above observation implies that $\ket{j,\bsy}$ is written in the form of
\eq{
\ket{j,\bsy}=\ket{\bsy'}\otimes (\otimes_{i=\abs{C}+1}^{L} V\ket{e_{k_i}}),
}
where $k_i\in \{1,2\}$.
This confirms that two states at different steps with the same initial state are orthogonal to each other:
\eq{
\braket{j,\bsy|j',\bsy}=0
}
for any $j\neq j'$.

\subsection{Case when TM2 halts 2: long-time average of $\calA$}\lb{s:halt2}

Using the eigenenergies and eigenstates shown in Eqs.~\eqref{exact-E} and \eqref{exact-psi}, $e^{-i\tH t}\ket{1,\bsy}$ is computed as
\balign{
e^{-i\tH t}\ket{1,\bsy}=&\sqrt{\frac{2}{\Jy+1}}\sum_{k=1}^{\Jy}e^{-iE_{k,\bsy} t}\sin \frac{k\pi}{\Jy+1}\ket{E_{k,\bsy}} \nt \\
=&\frac{2}{\Jy+1}\sum_{k=1}^{\Jy}e^{-iE_{k,\bsy} t}\sin \frac{k\pi}{\Jy+1} \sum_{j=1}^{\Jy} \sin \frac{jk\pi}{\Jy+1}\ket{j,\bsy},
}
where
\balign{
E_{k,\bsy}&:=2\cos \frac{k\pi}{\Jy+1}, \lb{Eky} \\
\ket{E_{k,\bsy}}&:=\sqrt{\frac{2}{\Jy+1}} \sum_{j=1}^{\Jy} \sin \frac{jk\pi}{\Jy+1}\ket{j,\bsy}
}
are the $k$-th energy eigenvalue and corresponding energy eigenstate of the effective Hamiltonian $\Heff=\sum_j \ket{j+1,\bsy}\bra{j,\bsy}+{\rm c.c.}$, and $\Jy$ is the total number of steps for the termination of $\tMG$ starting from the configuration in the cluster $C$ as $\bsy$ .

Using \eref{time-ave}, the long-time average of $\calA$ from the initial state $\ket{\psi_{V,L}^*}$ given in \eref{psip} is calculated as
\eqa{
\bA=\sum_{k,k', \bsy, \bsy'}\chi(E_{k,\bsy}=E_{k',\bsy'}) c^*_{\bsy} c_{\bsy'}\sqrt{\frac{2}{\Jy+1}}\sqrt{\frac{2}{\Jyp+1}}\sin \frac{k\pi}{\Jy+1}\sin \frac{k'\pi}{\Jyp+1}\braket{E_{k,\bsy}|\calA_L|E_{k',\bsy'}},
}{bA-start}
where $\chi(E_{k,\bsy}=E_{k',\bsy'})$ takes 1 if $E_{k,\bsy}=E_{k',\bsy'}$ and takes zero otherwise.
The contribution from the case of $\bsy=\bsy'$ (diagonal elements) has already been calculated and shown to be a finite amount in \sref{pf3}.
We shall prove that the contribution to $\bA$ from the case of $\bsy\neq \bsy'$ (off-diagonal elements) is sufficiently small.

Due to the form of \eref{Eky}, the condition $E_{k,\bsy}=E_{k',\bsy'}$ implies
\eqa{
\frac{k}{k'}=\frac{\Jy+1}{\Jyp+1}.
}{kkp}
Let $G$ be the greatest common divisor of $\Jy+1$ and $\Jyp+1$, and define $k_0:=\frac{\Jy+1}{G}$ and $k'_0:=\frac{\Jyp+1}{G}$.
Then, $k$ and $k'$ with \eref{kkp} are expressed as $k=lk_0$ and $k'=lk'_0$ with $l=1,2,\ldots , G$, and thus
\balign{
\frac{k}{\Jy+1}=\frac{k'}{\Jyp+1}=&\frac{l}{G}
}
is satisfied.
Hence, for any $\bsy\neq \bsy'$ we have
\balign{
&\sum_{\substack{k,k' \\ E_{k,\bsy}=E_{k',\bsy'}}}\sin \frac{k\pi}{\Jy+1}\sin \frac{k'\pi}{\Jyp+1}\braket{E_{k,\bsy}|\calA_L|E_{k',\bsy'}} \nt \\
=&\sum_{l=1}^{G-1} \sin^2 \frac{l\pi}{G}\braket{E_{k_0l,\bsy}|\calA_L|E_{k'_0l,\bsy'}} \nt \\
=&\sqrt{\frac{2}{\Jy+1}}\sqrt{\frac{2}{\Jyp+1}}  \sum_{l=1}^{G}\sum_{j=1}^{\Jy} \sum_{j'=1}^{\Jyp}\sin \frac{jl\pi}{G} \sin \frac{j'l\pi}{G} \sin^2 \frac{l\pi}{G}\braket{j,\bsy|\calA_L|j',\bsy'} \nt \\
=&\frac12 \sqrt{\frac{2}{\Jy+1}}\sqrt{\frac{2}{\Jyp+1}}  \sum_{l=1}^{G}\sum_{j=1}^{\Jy} \sum_{j'=1}^{\Jyp}\(\cos \frac{(j+j')l\pi}{G} +\cos \frac{(j-j')l\pi}{G}\) \sin^2 \frac{l\pi}{G} \braket{j,\bsy|\calA_L|j',\bsy'} \nt \\
\leq&\frac12 \sqrt{\frac{2}{\Jy+1}}\sqrt{\frac{2}{\Jyp+1}}  \sum_{l=1}^{G}\sum_{j=1}^{\Jy} \sum_{j'=1}^{\Jyp}\(\cos \frac{(j+j')l\pi}{G} +\cos \frac{(j-j')l\pi}{G}\)  \braket{j,\bsy|\calA_L|j',\bsy'} \nt \\
\leq&\frac12 \sqrt{\frac{2}{\Jy+1}}\sqrt{\frac{2}{\Jyp+1}}  (\Jy+1)(\Jyp+1) \braket{j,\bsy|\calA_L|j',\bsy'} \nt \\
\leq &\sqrt{(\Jy+1)(\Jyp+1)} \frac{\|A\|}{L}. \lb{sinsum}
}
In the sixth line, we used the fact that $\sum_{l=1}^G \cos \frac{(j+j')l\pi}{G}$ is equal to $G$ if $j+j'$ is a multiple of $2G$ and is equal to zero otherwise.
In the seventh line, we used the following relation similar to \eref{nonuni-off}
\eq{
\abs{\braket{j,\bsy|\calA_L|j',\bsy'}}=\abs{\frac{1}{L}\sum_{i=1}^{L}\braket{j,\bsy|A_i|j',\bsy'}} \leq \frac{\|A\|}{L}
}
for any $\bsy\neq \bsy'$.
As is the case of \eref{nonuni-off}, this relation follows from the fact that the configurations $(j,\bsy)$ and $(j',\bsy')$ differs at least a single site.

Substituting \eref{sinsum} into \eref{bA-start} in case of $\bsy\neq \bsy'$, we arrive at the upper bound for the off-diagonal sum:
\balign{
&\abs{\sum_{\substack{k,k', \bsy, \bsy' \\ \bsy\neq \bsy'}}\chi(E_{k,\bsy}=E_{k',\bsy'}) c^*_{\bsy} c_{\bsy'}\sqrt{\frac{2}{\Jy+1}}\sqrt{\frac{2}{\Jyp+1}}\sin \frac{k\pi}{\Jy+1}\sin \frac{k'\pi}{\Jyp+1}\braket{E_{k,\bsy}|\calA_L|E_{k',\bsy'}}} \nt \\
\leq & \frac{2\|A\|}{L}\sum_{ \bsy\neq \bsy'} \abs{c^*_{\bsy} c_{\bsy'}} \leq \frac{2\|A\|}{L} \(\sum_{\bsy} \abs{c_{\bsy}}\) ^2 \leq \frac{2\|A\|}{L}\abs{Y^*}. \lb{halt-off-bound}
}
Notably, $\abs{Y^*}$ is independent of the total system size $L$, and hence, by taking the thermodynamic limit $L\to \infty$ the right-hand side becomes arbitrarily small.
Thus, using the result in \sref{pf3}, we have
\eq{
\bA \geq \frac14 \braket{e_2|A|e_2}-\alpha \|A\| 
}
and 
\balign{
\overline{\calV\calA \calV^\dagger} \geq& \frac14 \braket{e_2|VA V^\dagger|e_2}-\alpha \|A\| \nt \\
=&\frac14 (\sqrt{\alpha}\ket{s_0, a_k}\bra{\psiin}+\sqrt{1-\alpha}\bra{e_k}) A  (\sqrt{\alpha}\ket{s_0, a_k}\ket{\psiin}+\sqrt{1-\alpha}\ket{e_k})-\alpha \|A\| \nt \\
\geq& \frac{1-\alpha}{4} \braket{e_2|A|e_2} -2(\alpha +\sqrt{\alpha (1-\alpha)})\|A\| .
}
By taking $\alpha$ sufficiently small, we arrive at the desired result 
\eq{
\min \{ \bA, \overline{V\calA V^\dagger}\} \geq \frac{1-\alpha}{4} \braket{e_2|A|e_2} -2(\alpha +\sqrt{\alpha (1-\alpha)})\|A\| \geq \(\frac14 -\eta\)  \braket{e_2|A|e_2}
}
for any $\eta>0$.

\subsection{Case when TM2 does not halt}\lb{s:nonhalt}

In this subsection, we consider the case that TM2 does not halt with the input $\bsu$.
To bound the off-diagonal elements from above, we need a completely different treatment from the case of halting, because the size of the cluster $C$ now becomes the entire system (The cluster $C$ should contain the working space of TM2, which is unlimited in case of non-halting).

To treat the non-halting case, we focus on the fact that most of the cells in the initial state are A-cells.
By construction, the number of A-cells is invariant under the time evolution, and if the decoding of input succeeds, all of A-cells are filled with the symbol $a_1$ at all times.

We first restrict the state space to ``good" configurations of $\bsx$.
In this subsection, we employ the word ``good" initial configurations with a slightly different definition from the previous subsection.
We say that $\bsx$ is a ``good" initial configuration if (i) the input for TM2 is correctly decoded by TM1, and (ii) the fraction of A-cells is larger than $1-\alpha'$ where $\alpha'$ is set properly as slightly larger than but close to $\alpha$.
We do not require the size of the working space for TM2.
We write as $X^*$ the set of good initial configurations in the above sense, and define the following unnormalized state:
\eqa{
\ket{\psi_{L}^{**}}:=\sum_{\bsx\in X^*}c_{\bsx} \ket{\bsx}.
}{psipp}
For a similar reason to the case of $Y^*$ in \sref{halt1}, the dynamics starting from $\ket{\psi_{L}^{**}}$ denoted by $\ket{\psi_{L}^{**}(t)}$ is arbitrarily close to the actual dynamics $\ket{\psi_{L}(t)}$.
Therefore, in the following, we consider the state $\ket{\psi_{L}^{**}(t)}$ instead of the actual one.

Let $P:=\ket{e_1}\bra{e_1}$ be the projector onto the A-cell filled with the symbol $a_1$, and $P_i$ be the aforementioned projector acting on the $i$-th site.
If $\ket{\bsx}$ represents a single configuration of $\tMG$, $\Tr[P_i \ket{\bsx}\bra{\bsx}]$ equals 1 (resp, 0) iff the $i$-th cell is (resp. is not) an A-cell with the symbol $a_1$.
Hence, we have
\eq{
\Tr \[ \sum_{i=1}^{L+1}P_i \ket{\psi_{L}^{**}(t)}\bra{\psi_{L}^{**}(t)}\] \geq (1-\alpha')L
}
for any $t$.
Let $\rho_i(t)$ denote the reduced density operator of $\ket{\psi_{L}^{**}(t)}\bra{\psi_{L}^{**}(t)}$ onto the $i$-th site, and define the averaged density operator over all sites as
\eq{
\orho(t):=\frac{1}{L}\sum_{i=1}^{L}\rho_i(t).
}
Then, using a relation
\eq{
\Tr[P\orho(t)]=\braket{e_1|\orho(t)|e_1}\geq 1-\alpha',
}
we arrive at the desired result:
\balign{
\abs{\braket{\psi_{L}(t)|\calA_L|\psi_{L}(t)}}=&\abs{\braket{\psi_{L}(t)|\calA_L|\psi_{L}(t)}-\braket{e_1|A|e_1}} \nt \\
=&\abs{\Tr[A\orho(t)]-\braket{e_1|A|e_1}} \nt \\
\leq& \| \orho(t)-\ket{e_1}\bra{e_1}\|_1 \|A\| \nt \\
\leq& 2\sqrt{1-\frac{(1-\alpha')L}{L+1}}\|A\| \nt \\
<& 2\sqrt{\alpha'}\|A\| . \lb{main-nonhalt}
}
Since we can set $\alpha'<2\alpha$, we arrive at the desired relation $\bA<2\sqrt{2\alpha}\|A\|$.

To show Lemma 1, we also need to evaluate the state with $V$ rotation, which is evaluated as
\balign{
&\abs{\braket{\psi_L|\calV \calA_L \calV^{\dagger}|\psi_L}-\braket{e_1|A|e_1}} \nt \\
\leq&\abs{\braket{\psi_L|\calV \calA_L \calV^{\dagger}|\psi_L}-\braket{e_1|VA V^\dagger|e_1}}+\abs{\braket{e_1|VA V^\dagger|e_1}-\braket{e_1|A|e_1}}.
}
The first term has readily obtained through \eref{main-nonhalt} with replacing $A$ to $VAV^\dagger$, which bounds the first term from above by $2\sqrt{2\alpha}\|A\|$.
The second term is calculated as
\balign{
\abs{\braket{e_1|VA V^\dagger|e_1}-\braket{e_1|A|e_1}}
\leq &\|A\| \| \ket{e_1}\bra{e_1}-V^\dagger \ket{e_1}\bra{e_1}V\| \nt \\
=&\|A\| \sqrt{1-\braket{e_1|V|e_1}} \nt \\
=&\|A\| \sqrt{\alpha},
}
where in the last line we used Eqs.~\eqref{V-def} and \eqref{ek-def}.
Combining them, we find
\eq{
\max \{ \bA, \overline{V\calA V^\dagger}\} \leq 4\sqrt{\alpha}\|A\| ,
}
whose right-hand side can become arbitrarily small by taking $\alpha$ sufficiently small.
In particular, $\max \{ \bA, \overline{V\calA V^\dagger}\} \leq \eta$ is fulfilled, which completes the whole proof of Lemma 1.

\subsection{Note: dimension of the local Hilbert space}\lb{s:dim}

We have not discussed how large the sufficient dimension of the local Hilbert space is.
We here present a rough estimation of it instead of rigorous calculation.

To reduce the dimension, it is useful to enlarge the tape alphabet.
More symbols the tape alphabet has, fewer states the finite control needs to have.
It has been established that there are URTMs with 10-state and 8-symbol, 15-state and 6-symbol, 24-state and 4-symbol, 32-state and 3-symbol, and 138-state and 2-symbol~\cite{Morbook}.
By employing the 24-state 4-symbol URTM, 24 states in the finite control and $(4+1)\times 2\times 2^2=40$ symbols in the tape alphabet suffice to simulate TM2 (see also \fref{state}).
Although we do not evaluate the sufficient number of states and symbols for TM1, since the tasks of TM1 are all elementary; counting the number of 1's and 0's. computing logarithm, and division, we roughly estimate that 50 states and 4 symbols suffice for TM1.
If the above estimation is correct, the sufficient dimension of the local Hilbert space is $(24+50+1)+(40+2+2)=119$.

\section{Proof of Theorem 2 (undecidability of thermalization)}\lb{s:proof-thermal}

Our main idea to prove Theorem 2 is tuning the equilibrium value $\Tr[\calA\rho^{\rm MC}]$ to the desired target value $A^*$ by changing the Hamiltonian.
To this end, we need some refinements and modifications of the proof of Lemma 1.
In Sec.\ref{s:bA-large}, we first specify the long-time average of $\calA$ when the URTM halts.
In Sec.\ref{s:bA-TM3+}, we modify the dynamics of TM3, which enables us to push the value of $\bA$ in case of halting away from that in case of non-halting.
In Sec.\ref{s:tune-thermo}, we construct a family of Hamiltonians whose microcanonical average of $\calA$ is tuned to the target value $A^*$.

\subsection{Value of $\bA$ in thermodynamic limit when TM2 halts}\lb{s:bA-large}

In the previous section, we only derive an inequality for the value of $\bA$ when TM2 halts.
We here briefly calculate the value of $\bA$ in the thermodynamic limit when TM2 halts.

We start from \eref{nonuni-correction} and the equality part of \eref{time-ave-nonuni-fin} for non-uniform (computational basis) initial states.
By taking $L$ to infinity, the number of steps before TM2 halts is negligibly small compared to the total number of steps $J$.
Thus the right-hand side in the equality of \eref{time-ave-nonuni-fin} is calculated as
\eq{
\lim_{L\to \infty}\frac{3}{2(J+1)}\( \braket{\bsx^1|\calA_L|\bsx^1}+\braket{\bsx^J|\calA_L|\bsx^J}\) +\frac{1}{J+1}\sum_{j=2}^{J-1}\braket{\bsx^j|\calA_L|\bsx^j} =\frac{1-\alpha}{2}\braket{e_2|A|e_2}
}
for computational basis initial states, where we assumed that M-cells and A-cells are uniformly distributed.
By combining \eref{nonuni-correction}, this relation leads to the bound for $\bA$ for computational basis initial states:$\abs{\bA -\frac{1-\alpha}{2}\braket{e_2|A|e_2}}\leq \alpha \|A\|$.
In the case of uniform initial states (excepts for the first site), we have already shown that off-diagonal terms and various other correction terms vanish in the thermodynamic limit.
Thus, by using Eqs.~\eqref{bA-start} and \eqref{halt-off-bound}, we readily have
\balign{
\abs{\bA -\frac{1-\alpha}{2}\braket{e_2|A|e_2}}&\leq \alpha \|A\| , \\
\abs{\overline{V\calA V^\dagger} - \frac{1-\alpha}{2} \braket{e_2|A|e_2}} &\leq 2(\alpha +\sqrt{\alpha (1-\alpha)})\|A\| ,
}
for uniform initial states.
This means that both $\bA$ and $\overline{V\calA V^\dagger}$ converge close to $\frac{1-\alpha}{2}\braket{e_2|A|e_2}$ in the thermodynamic limit with errors $\alpha \|A\|$ and $2(\alpha +\sqrt{\alpha (1-\alpha)})\|A\|$.
Since $\alpha$ can be taken arbitrarily small, these relations roughly state that both $\bA$ and $\overline{V\calA V^\dagger}$ converge around $\frac12 \braket{e_2|A|e_2}$ with arbitrarily small errors.

\subsection{Pushing the value of $\bA$ away from zero}\lb{s:bA-TM3+}

\figin{15cm}{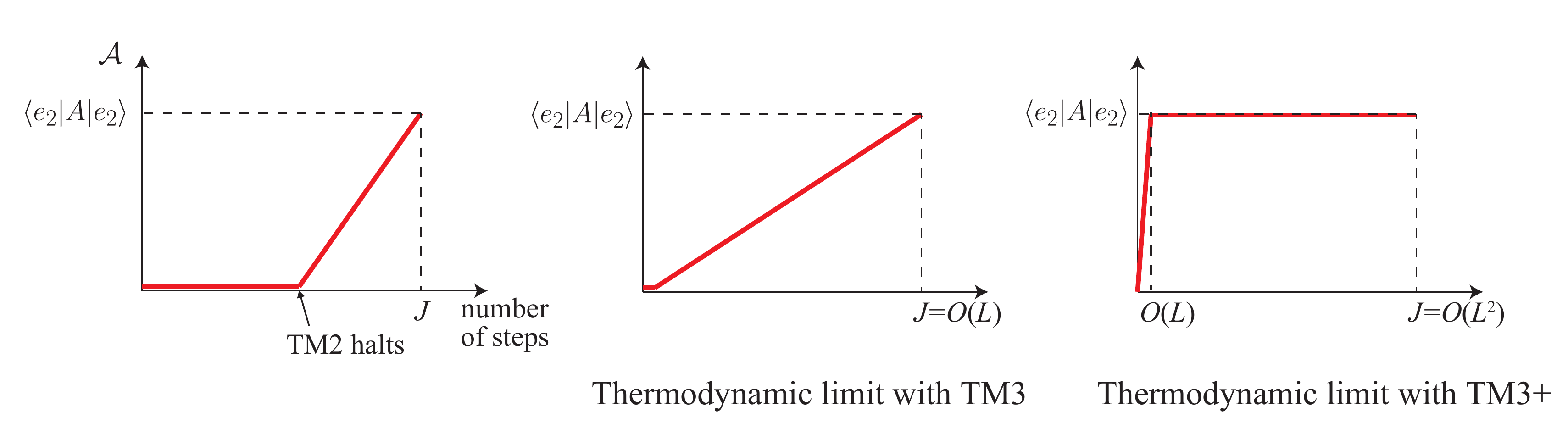}{
(Left): A typical graph of the behavior of the expectation value of $A$ inf the classical TM (with TM1, TM2, and TM3) for the number of steps.
After TM2 halts, the expectation value of $A$ starts to increase and reaches $\braket{e_2|A|e_2}$ at the final step $J$.
(Middle): Thermodynamic limit of the left graph.
Since TM2 halts in constant steps, in $L\to \infty$ limit the dynamics of TM3 dominates this graph.
(Right): Thermodynamics limit when we use TM3+ instead of TM3.
After flipping all A-cells from $a_1$ to $a_2$, TM3+ runs for around $L^2/2$ steps, which makes the long-time average $\bA$ arbitrarily close to $\braket{e_2|A|e_2}$.
}{A-graph}

In the setup in the proof of Lemma 1, the value of $\bA$ when TM2 halts is shown to approach $\frac12 \braket{e_2|A|e_2}$.
This value, more precisely $\frac12 (\braket{e_1|A|e_1}+ \braket{e_2|A|e_2})$, comes from the following calculation:
In the thermodynamic limit, the time evolution of $A(t)$ in the classical TM is roughly described by a linear function in time, and the dynamics stops when $A(t)$ reaches $\braket{e_2|A|e_2}$ (see the middle graph in \fref{A-graph}).
Thus, the time average of $\calA(t)$ is $\frac12 \braket{e_2|A|e_2}$.

In order to inflate $\bA$, we modify the rule of TM3 such that the machine spends most of the time with $\calA(t)$ at $\braket{e_2|A|e_2}$.
The modified TM3, named TM3+, has two additional states in A-cells, and three additional internal states in the finite control.
We first recall the states and the rule of TM3\fn{
Since TM3 passes M-cells, we omit the description on M-cells.
}:
In TM3, A-cells take two different states, $a_1$ and $a_2$, and the finite control has a single internal state $r$.
The rule of the dynamics is as follows:
\bi{
\item If the head reads $a_1$ with the internal state $r$, then the state of the A-cell is changed to $a_2$ and the finite control moves right.
\item If the head reads $a_2$ with the internal state $r$, then the finite control stops.
}

\figin{9cm}{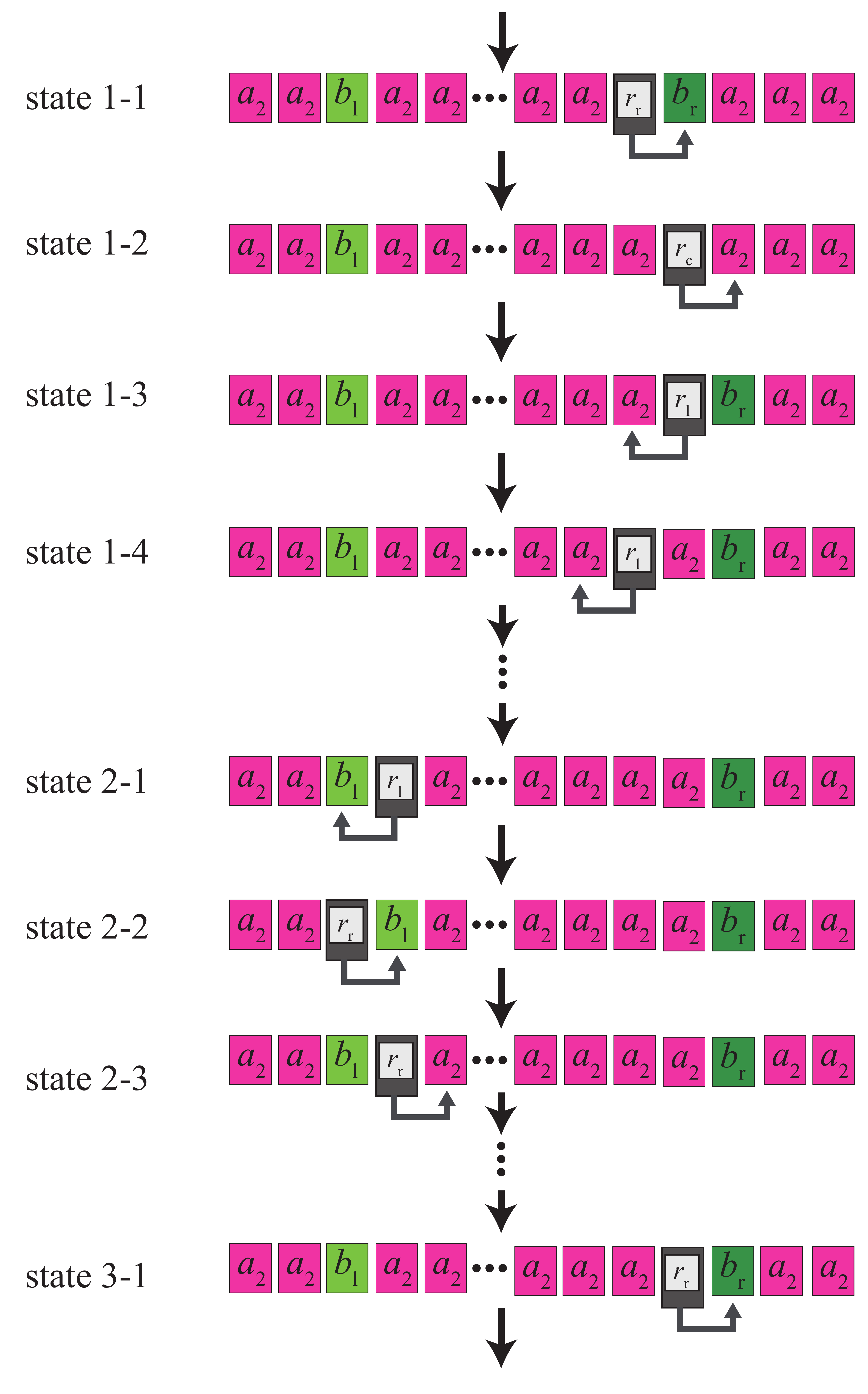}{
Schematic of the dynamics of TM3+.
We drop M-cells for brevity.
The states 1-1 and 3-1 are very similar, but the cell with $\br$ moves right one cell.
}{full-TM3+}

Now, we describe the states and the rule of TM3+.
In TM3+, A-cells take four different states, $a_1$, $a_2$, $\bl$ and $\br$, and the finite control has four different internal states, $r$, $\rl$, $\rr$, and $\rc$.
The rule of the dynamics is as follows:
\bi{
\item If the head reads $a_1$ with the internal state $r$, then the state of the A-cell is changed to $a_2$, and the finite control moves right.
\item If the head reads $a_2$ with the internal state $r$, then the state of the A-cell is changed to $\bl$, the internal state is changed to $\rc$, and the finite control moves right.
\item If the head reads $a_2$ with the internal state $\rc$, then the state of the A-cell is changed to $\br$, the internal state is changed to $\rl$, and the finite control does not move (state 1-2 to 1-3 in \fref{full-TM3+}).
\item If the head reads $a_2$ with the internal state $\rl$, then the state of the A-cell and the internal state are kept unchanged, and the finite control moves left (state 1-3 to1-4).
\item If the head reads $\bl$ with the internal state $\rl$, then the state of the A-cell is kept unchanged, the internal state is changed to $\rr$, and the finite control moves left (state 2-1 to 2-2).
\item If the head reads $a_2$ or $\bl$ with the internal state $\rr$, then the state of the A-cell and the internal state is kept unchanged, and the finite control moves right (state 2-2 to 2-3 and 2-3 to 3-1).
\item If the head reads $\br$ with the internal state $\rr$, then the state of the A-cell is changed to $a_2$, the internal state is changed to $\rc$, and the finite control moves right (state 1-1 to 1-2).
\item If the head reads $\bl$ with the internal state $\rc$, then the finite control stops.
}
This rule yields the following dynamics:
After flipping all A-cells to $a_2$, then TM3+ shuttles between $\bl$ and $\br$ with pushing $\br$.
AT the beginning of this shuttle, the cell $\br$ sits the right of $\bl$.
During a single shuttle between $\bl$ and $\br$, $\bl$ does not move on average and $\br$ moves right one cell (see \fref{full-TM3+}).
This shuttle finishes when $\br$ hits $\bl$ from the left.
The flipping from $a_1$ to $a_2$ takes $O(L)$ steps, while the shuttle process after flipping takes $O(L^2)$ steps.
Namely, at almost all times, all A-cells take the state $a_2$.
Thus, following the argument in the previous subsection, we find that in the corresponding quantum system the value of $\bA$ approaches $\braket{e_2|A|e_2}$ when TM2 halts.

\subsection{Tuning the microcanonical average}\lb{s:tune-thermo}

We now demonstrate how to tune the long-time average $\bA$ in case of halting to the microcanonical average $\Tr[\calA\rho^{\rm MC}]$.
As seen in the previous subsections, by employing TM3+ the long-time average $\bA$ in case of halting (resp. non-halting) approaches $\braket{e_2|A|e_2}$ (resp. $\braket{e_1|A|e_1}=0$) in the thermodynamic limit.
We shall show the existence of a proper orthonormal set\fn{
This set is not CONS because this set does not span the whole Hilbert space.
However, since the dynamics are closed in this subspace spanned by this set, this point does not matter to our argument.
(We assign extremely large energy to states outside this subspace in order to keep the contribution from these states to microcanonical average negligible).
}
$\{ \ket{e_i}\}$ where $\Tr[\calA\rho^{\rm MC}]=\braket{e_2|A|e_2}$ holds.

We introduce the Hilbert subspace with states orthogonal to $\ket{\phi_0}$, $\ket{\phi_1}$, $A\ket{\phi_0}$ and $A\ket{\phi_1}$ denoted by $\calS\subset \calH$.
Let $P_{\calS}$ be the projection operator onto $\calS$.
We consider an operator $P_{\calS}AP_{\calS}$ on $\calS$, and express its eigenstates as $\{ \ket{\sigma_i}\}$.
Since $P_{\calS}\ket{\sigma_i}=\ket{\sigma_i}$ and $P_{\calS}\ket{\sigma_j}=\ket{\sigma_j}$, we have
\eq{
\braket{\sigma_i|A|\sigma_j}=0 \hspace{10pt} (i\neq j).
}
We set the indices of $\{ \ket{\sigma_i}\}$ in decreasing order of expectation value of $A$;
\eq{
\braket{\sigma_1|A|\sigma_1}\geq \braket{\sigma_2|A|\sigma_2}\geq \cdots \geq \braket{\sigma_{d'}|A|\sigma_{d'}},
}
where we denoted the dimension of $\calS$ by $d'$.
Thanks to the assumption $\braket{\phi_2|A|\phi_2}>\max_{\ket{\psi}\in {\rm span}\{ \ket{\phi_0}, \ket{\phi_1}\}} \braket{\psi|A|\psi}$ and $\braket{\phi_3|A|\phi_3}<\min_{\ket{\psi}\in {\rm span}\{ \ket{\phi_0}, \ket{\phi_1}\}} \braket{\psi|A|\psi}$, we find\fn{
Since $\ket{\phi_2}\in \calS$, we can expand $\ket{\phi_2}$ as $\ket{\phi_2}=\sum_i c_i \ket{\sigma_i}$.
Due to the diagonal property of $A$ with the orthonormal set $\{\ket{\sigma_i}\}$, $\braket{\phi_2|A|\phi_2}=\sum_i \abs{c_i}^2 \braket{\sigma_i|A|\sigma_i}<\braket{\sigma_1|A|\sigma_1}$.
A similar argument holds for $\ket{\phi_3}$.
}
\balign{
\braket{\sigma_1|A|\sigma_1}&>\max_{\ket{\psi}\in {\rm span}\{ \ket{\phi_0}, \ket{\phi_1}\}} \braket{\psi|A|\psi}, \\
\braket{\sigma_{d'}|A|\sigma_{d'}}&<\min_{\ket{\psi}\in {\rm span}\{ \ket{\phi_0}, \ket{\phi_1}\}} \braket{\psi|A|\psi}.
}

Let $\rho^{\rm MC}_1$ be the reduced density matrix of a microcanonical state to a single site with energy $\braket{\psi_0|H|\psi_0}$ ($\ket{\psi_0}:=\ket{\phi_0}\otimes \ket{\phi_1}\otimes \cdots \otimes \ket{\phi_1}$).
Due to the translation invariance of $H$, $\Tr[\calA\rho^{\rm MC}]=\Tr[A\rho_1^{\rm MC}]$ is satisfied.
Note that the Hamiltonian $H$ depends on the input code $\bsu$ and several parameters, and thus  $\rho^{\rm MC}_1$ also depends on them.
We formally expand it with the set $\{\ket{e_i}\}$ as $\rho^{\rm MC}_1=\sum_{i,j}\ket{e_i}\bra{e_j}\braket{e_i|\rho^{\rm MC}_1|e_j}=:\sum_{i,j}\ket{e_i}\bra{e_j}\rho^{\rm MC}_{ij}$.
We have already set $\ket{e_0}=\ket{\phi_0}$ and $\ket{e_1}=\ket{\phi_1}$.

We first consider the case that $\braket{\sigma_2|A|\sigma_2}\neq \braket{\sigma_{d'-1}|A|\sigma_{d'-1}}$.
We set the orthonormal set $\{\ket{e_i}\}$ as
\balign{
\ket{e_2}&=\sqrt{p}\ket{\sigma_1}+\sqrt{1-p}\ket{\sigma_{d'}}, \\
\ket{e_3}&=\sqrt{q}\ket{\sigma_2}+\sqrt{1-q}\ket{\sigma_{d'-1}}, \\
\ket{e_i}&=\ket{\sigma_{i-1}},
}
with $i\geq 4$.
By permuting the label of $\{\ket{e_i}\}$ if necessary, without loss of generality we suppose that the diagonal elements of $\ket{e_3}$ in the reduced microcanonical state is nonzero: $\rho^{\rm MC}_{33}\neq 0$.
Since all off-diagonal elements of $A$ with this orthonormal set is zero, the microcanonical average reads
\eq{
\Tr[A\rho^{\rm MC}_1]=\sum_{i=2}\braket{e_i|A|e_i}\rho^{\rm MC}_{ii}+\braket{e'|A|e'}\braket{e'|\rho^{\rm MC}_1|e'}+\braket{e''|A|e''}\braket{e''|\rho^{\rm MC}_1|e''},
}
where $\ket{e'}, \ket{e''}\in \calT:= {\rm span}\{ \ket{\phi_0}, \ket{\phi_1}\}$ are proper states diagonalizing $P_{\calT}AP_{\calT}$ in this subspace $\calT$.
By construction, for any $p$ and $q$,
\eq{
\braket{\sigma_1|A|\sigma_1}> \Tr[A\rho^{\rm MC}_1] >\braket{\sigma_{d'}|A|\sigma_{d'}}
}
is satisfied.

\figin{7cm}{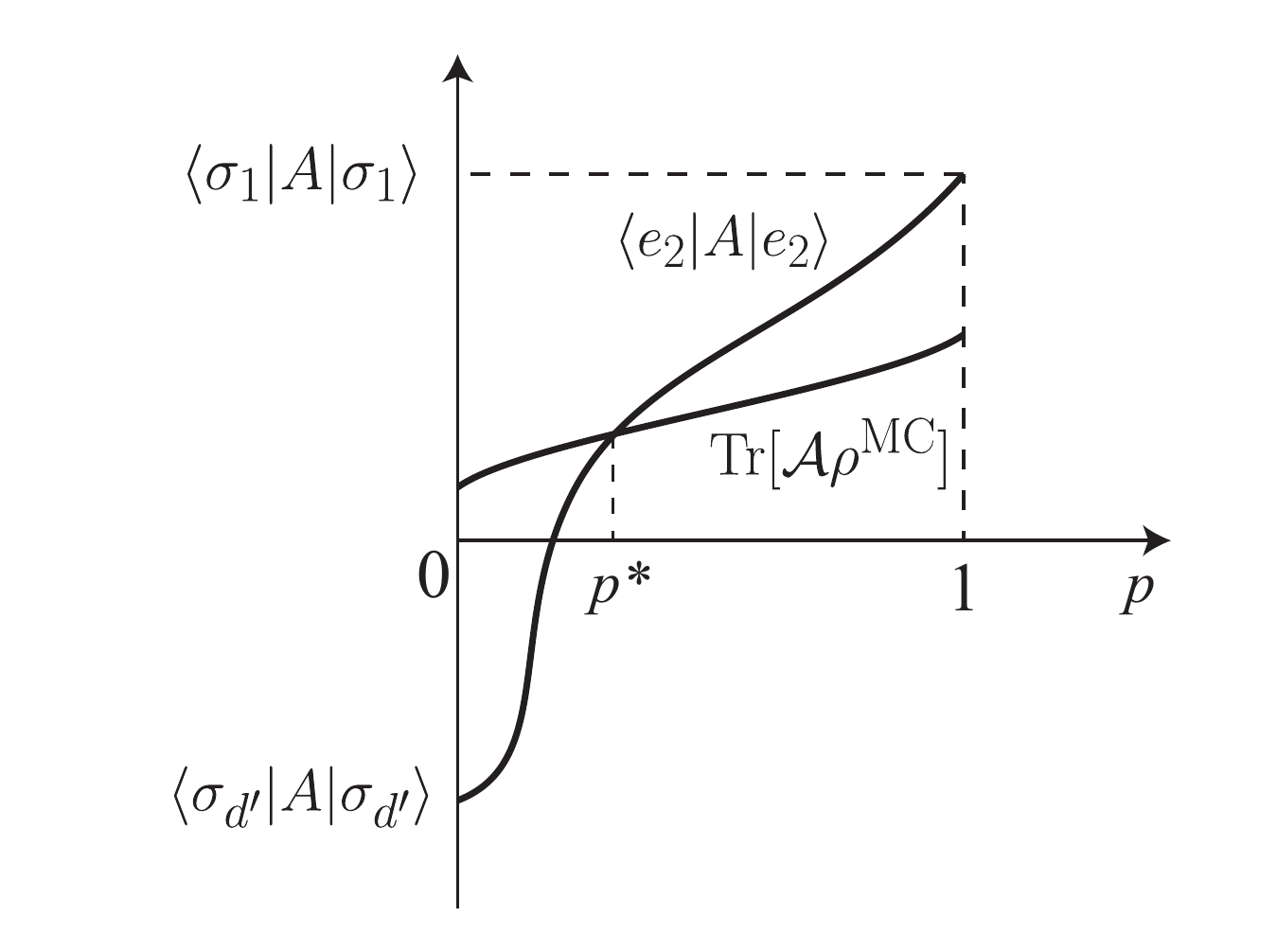}{
Tuning of $p$ from $p=0$ to $p=1$.
Owing to $\braket{\sigma_1|A|\sigma_1}> \Tr[A\rho^{\rm MC}_1] >\braket{\sigma_{d'}|A|\sigma_{d'}}$, there must exist a proper $p$ (denoted by $p^*$ in this graph) with which $\braket{e_2|A|e_2}=\Tr[A\rho^{\rm MC}_1]$ is satisfied.
}{full-tune}

We now change the parameter $p$ from $p=0$ to $p=1$ continuously with a fixed $q$.
Accordingly the long-time average $\braket{e_2|A|e_2}$ changes from $\braket{\sigma_{d'}|A|\sigma_{d'}}$ to $\braket{\sigma_1|A|\sigma_1}$ continuously.
Hence, there exists a proper $0\leq p^* \leq 1$ which fulfills $\braket{e_2|A|e_2}=\Tr[A\rho^{\rm MC}_1]$ (see \fref{full-tune}).
Note that by tuning $q$ if needed, we can safely avoid the undesired situation that the above coincidence happens at $\Tr[A\rho^{\rm MC}_1]=\braket{e_2|A|e_2}=0$.
In summary, by employing the above $p^*$, these orthonormal set $\{ \ket{e_i}\}$ realizes the Hamiltonian with which the long-time average $\bA$ is equal to $\Tr[A\rho^{\rm MC}_1]\neq 0$ if and only if the URTM with the input $\bsu$ halts.

We next consider the case that $\braket{\sigma_2|A|\sigma_2}=\braket{\sigma_3|A|\sigma_3}=\cdots = \braket{\sigma_{d'-1}|A|\sigma_{d'-1}}\neq \braket{\sigma_1|A|\sigma_1}$
\fn{
In case of $\braket{\sigma_1|A|\sigma_1}=\braket{\sigma_2|A|\sigma_2}=\cdots = \braket{\sigma_{d'-2}|A|\sigma_{d'-2}}$, by reversing the role of $\ket{\sigma_1}$ and $\ket{\sigma_{d'}}$ a similar argument holds.
}.
In this case, we set
\balign{
\ket{e_2}&=\sqrt{p}(\sqrt{q}\ket{\sigma_1}-\sqrt{1-q}\ket{\sigma_2})+\sqrt{1-p}\ket{\sigma_{d'}} \\
\ket{e_3}&=\sqrt{1-q}\ket{\sigma_1}+\sqrt{q}\ket{\sigma_2} \\
\ket{e_i}&=\ket{\sigma_{i-1}}
}
with $i\geq 4$, where we again supposed $\rho^{\rm MC}_{33}\neq 0$ without loss of generality.
If $q$ is close to 1, we have $\braket{e_2|A|e_2}>\Tr[A\rho^{\rm MC}_1]$ by setting $p=1$.
Thus, by changing $p$ from $p=0$ to $p=1$, we again find that there exists a proper $p^*$ such that $\braket{e_2|A|e_2}=\Tr[A\rho^{\rm MC}_1]$ is satisfied.
We can avoid the unwanted situation $\Tr[A\rho^{\rm MC}_1]=\braket{e_2|A|e_2}=0$ by tuning $q$.
In summary, by employing the above $p^*$, these orthonormal set $\{ \ket{e_i}\}$ realizes the Hamiltonian with which the long-time average $\bA$ is equal to $\Tr[A\rho^{\rm MC}_1]\neq 0$ if and only if the URTM with the input $\bsu$ halts.
This completes the proof of Theorem 2.

\section{Remarks on the infinite limit}\lb{s:remark}

\subsection{What happens if we numerically simulate this system?}

Our result shows that the long-time average of $\calA$ is uncomputable even with the unlimited computational resource.
Therefore, one may wonder about the relation between the uncomputability and the fact that we can numerically simulate any system with finite size.
We here comment on this point.

\figin{12cm}{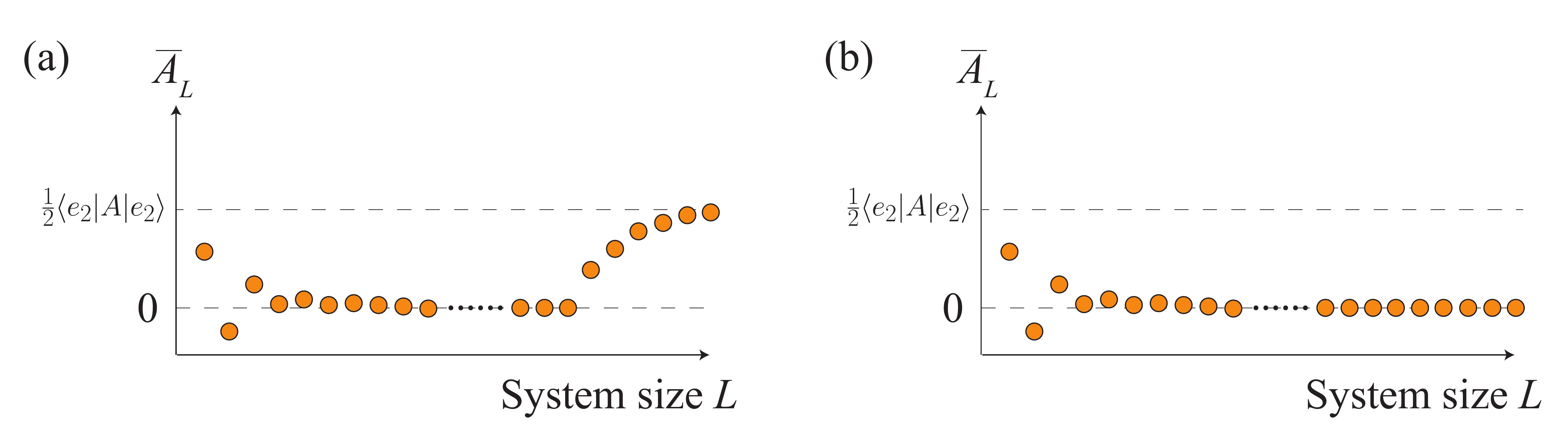}{
A possible situation in numerical exact diagonalization.
Although the long-time average $\bA_L$ appears to converge to some value, it may suddenly deviate from this value and converge to another value as (a).
We cannot distinguish this long metastable case (a) from a true convergence case (b).
}{diag}

Although there are several possible scenarios, we here describe the most intuitive one.
We consider the case that TM2 halts after very long steps.
If the system size is insufficient, the finite control passes the periodic boundary before halting and stops its movement.
With such a small system size, $\bA_L$ takes a value close to zero.
On the other hand, if the system size becomes sufficiently large for TM2 to halt, then $\bA_L$ start approaching around $\frac12 \braket{e_2|A|e_2}$.
The problem lies in the fact that we never know what system size is {\it sufficiently large}.
The undecidability of Turing machines tells that we cannot exclude the possibility that a Turing machine runs for very long steps and suddenly halts.
In terms of numerical simulations, any large numerical simulation might be fooled due to insufficient system size, and thus even with the unlimited computational resource we cannot distinguish the case with a long metastable state (\fref{diag}~(a)) and the case that the system has already relaxed (\fref{diag}~(b)).
This shows clear contrast to the numerical simulation of conventional many-body systems, where by taking sufficiently large system size we can make the amount of error from the true value in the thermodynamic limit arbitrarily small.

\subsection{Difference from the behavior of near-integrable systems}

The above explanation may convey the impression that our constructed system is essentially the same as near-integrable systems; $H=H_{\rm int}+\ep V$, with an integrable Hamiltonian $H_{\rm int}$ and a small perturbation $\ep V$.
In near-integrable systems, the small parameter $\ep$ characterizes the necessary system size to reach the true long-time average, which can become extremely large by taking $\ep$ close to zero.
By preparing a sufficiently large system, where its sufficiency is characterized by $\ep$, we succeed in reaching the correct long-time average within an arbitrarily small error.

We claim that our system does not have such a small parameter going to an arbitrarily small number, and as its consequence sufficiently large system size does not exist.
One may feel that the length of an input code $\bsu$, which induces the change in the input Hamiltonian $H$, serves as a small parameter.
If this guess is true, by preparing a large system whose size is determined by the input length, we safely observe the presence or absence of thermalization in this large system.
However, unfortunately, this is not the case.

To explain why, we introduce the {\it busy beaver function} BB$(n)$, which is the maximum number of steps taken by a {\it halting} Turing machine (TM) with $n$ different internal states and an empty input code.
We remark that a TM with $n$ internal states and an empty input can be implemented by a fixed URTM with a small number of internal states and an input code with length at most $l(n)$, and vice versa.
Thus, the busy beaver function can be regarded as the maximum number of steps the URTM can take with a halting input code with a given length.
This fact directly implies that the estimation of the sufficient size for thermalization with a given input code $\bsu$ is essentially the same as the computation of the busy beaver function.

However, the busy beaver function is proven not to be a computable function.
More surprisingly, it is also proven that we cannot compute the busy beaver function BB$(n)$ with $n\geq 748$ as far as the Zermelo-Fraenkel set theory with the axiom of choice (ZFC), which is roughly equivalent to our mathematics, is consistent~\cite{YA16, Aar-online}.
This striking result comes from the fact that there exists a 748-state TM which halts if and only if the ZFC is inconsistent.
The G\"{o}del's incompleteness theorem reveals that the ZFC cannot prove the consistency of the ZFC itself.
Hence, if BB$(748)$ is computable, then by running the above TM for ZFC consistency until the step determined by BB$(748)$ we can prove the consistency of the ZFC, which is a contradiction.

The incomputability of BB$(748)$ leads to the absence of small parameters in our system of thermalization.
We collect all Hamiltonians which correspond to the TMs with 748 internal states and empty input codes.
Since there are only a finite number of Hamiltonians, any two Hamiltonians in this set have a finite difference between them (i.e., there is no small parameter going to zero).
However, as we have seen, we do not have {\it sufficiently large system size} in spite of no small parameter in these Hamiltonians, as long as our mathematics is consistent.

\FS{
\section{Concluding remarks}
We have shown that in one-dimensional shift-invariant quantum many-body systems, whether the long-time average $\bA$ is in the vicinity of $c$ or not is undecidable.
The main idea of our proof is a map from a classical URTM to the Hamiltonian of quantum many-body systems.
To make the interaction as nearest-neighbor and shift-invariant, we employ the Feynman-Kitaev type construction of a Hamiltonian.

Our method to implement the halting problem to dynamics of quantum many-body systems will apply to various other problems related to thermalization.
However, to encode the halting problem, two cases, the halting case and the non-halting case, should be mapped onto two distinct situations (e.g., thermalizing in one case and not thermalizing in the other case).
Hence, the presence of both two phenomena should be proved in some concrete systems.
This prevents the extension of our method to the eigenstate thermalization hypothesis because no concrete model is proven to satisfy the eigenstate thermalization hypothesis.
In contrast, a recent study proves the non-integrability of some concrete models in the sense that the system has no local conserved quantity~\cite{S18}.
Therefore, the decision problem on (non-)integrability for a given Hamiltonian may serve as a stage to extend our method.

We remark that our result does not exclude the possibility that a particular model is proven to thermalize or not to thermalize.
Our result excludes the possibility to obtain a general and ultimate criterion to judge the presence or absence of thermalization, but not a model-specific result.
In fact, integrable systems are known not to thermalize.
We also remark that our model is highly artificial, which is a limitation of our result.
We thus have two interesting open problems:
One is to find a more {\it natural} model exhibiting undecidability.
The other is to construct a set of a more restricted class of Hamiltonians whose fate of thermalization is decidable.
Both problems merit further research.
}{}

\FS{
\section*{Acknowledgment}
We thank Takahiro Sagawa for stimulating discussion.
NS was supported by JSPS Grants-in-Aid for Scientific Research Grant Number JP19K14615. 
}{}

\end{document}